\pdfoutput=1
\documentclass[12pt,a4paper]{article}
\usepackage{ifthen} % for conditional statements
\newboolean{pdflatex}
\setboolean{pdflatex}{true} % False for eps figures 

\newboolean{articletitles}
\setboolean{articletitles}{true} % False removes titles in references

\newboolean{uprightparticles}
\setboolean{uprightparticles}{false} %True for upright particle symbols

\def\paperauthors{LHCb collaboration} % Leave as is for PAPER, CONF and FIGURE
\def\paperasciititle{First evidence of the Bs2KpiG decay} % Set ASCII title here !! MAKE sure it's only ASCII characters !! 
\def\papertitle{First evidence of the \decay{\Bs}{\Km\pip\gamma} decay} % Latex formatted title
\def\paperkeywords{{High Energy Physics}, {LHCb}} % Comma separated list
\def\papercopyright{\the\year\ CERN for the benefit of the LHCb collaboration} % new since 9/Apr/2018
\def\paperlicence{CC BY 4.0 licence}
\def\paperlicenceurl{https://creativecommons.org/licenses/by/4.0/}

\newif\ifEnableSectionTOCLinks
\EnableSectionTOCLinksfalse % deactivated

\usepackage[top=1in, bottom=1.25in, left=1in, right=1in]{geometry}

\usepackage{microtype}
\usepackage{lineno}  % for line numbering during review
\usepackage{xspace} % To avoid problems with missing or double spaces after
\usepackage{caption} %these three command get the figure and table captions automatically small

\usepackage{graphicx}  % to include figures (can also use other packages)
\usepackage{color}
\usepackage{colortbl}
\graphicspath{{./figs/}} % Make Latex search fig subdir for figures
\usepackage{amsmath} % Adds a large collection of math symbols
\usepackage{amssymb}
\usepackage{amsfonts}
\usepackage{upgreek} % Adds in support for greek letters in roman typeset

\newcommand*\patchAmsMathEnvironmentForLineno[1]{%
\expandafter\let\csname old#1\expandafter\endcsname\csname #1\endcsname
\expandafter\let\csname oldend#1\expandafter\endcsname\csname
end#1\endcsname
 \renewenvironment{#1}%
   {\linenomath\csname old#1\endcsname}%
   {\csname oldend#1\endcsname\endlinenomath}%
}
\newcommand*\patchBothAmsMathEnvironmentsForLineno[1]{%
  \patchAmsMathEnvironmentForLineno{#1}%
  \patchAmsMathEnvironmentForLineno{#1*}%
}
\AtBeginDocument{%
\patchBothAmsMathEnvironmentsForLineno{equation}%
\patchBothAmsMathEnvironmentsForLineno{align}%
\patchBothAmsMathEnvironmentsForLineno{flalign}%
\patchBothAmsMathEnvironmentsForLineno{alignat}%
\patchBothAmsMathEnvironmentsForLineno{gather}%
\patchBothAmsMathEnvironmentsForLineno{multline}%
\patchBothAmsMathEnvironmentsForLineno{eqnarray}%
}

\usepackage[pdftex,
            pdfauthor={\paperauthors},
            pdftitle={\paperasciititle},
            pdfkeywords={\paperkeywords}]{hyperref}
\usepackage{hyperxmp}
\hypersetup{
    pdfcopyright={Copyright (C) \papercopyright},
    pdflicenseurl={\paperlicenceurl}
}
\usepackage[colorinlistoftodos,textsize=scriptsize]{todonotes}

\usepackage[bottom,flushmargin,hang,multiple]{footmisc}

\usepackage[all]{hypcap} % Internal hyperlinks to floats.

\usepackage{xspace} 
\usepackage{upgreek}

\def\lhcb   {\mbox{LHCb}\xspace}

\def\MagUp {\mbox{\em Mag\kern -0.05em Up}\xspace}

\ifthenelse{\boolean{uprightparticles}}%
{

 \def\Ppi         {\ensuremath{\uppi}\xspace}                 
                  
 \def\Prho        {\ensuremath{\uprho}\xspace}

 \def\PDelta      {\ensuremath{\Delta}\xspace}                 
 \def\PXi         {\ensuremath{\Xi}\xspace}                 
 \def\PLambda     {\ensuremath{\Lambda}\xspace}                 
 \def\PSigma      {\ensuremath{\Sigma}\xspace}                 
 \def\POmega      {\ensuremath{\Omega}\xspace}                 
 \def\PUpsilon    {\ensuremath{\Upsilon}\xspace}
 \let\oldPi\Pi
 \def\PPi         {\ensuremath{\oldPi}\xspace}

 \def\PB      {\ensuremath{\mathrm{B}}\xspace}                 
 \def\PD      {\ensuremath{\mathrm{D}}\xspace}                 

 \def\PK      {\ensuremath{\mathrm{K}}\xspace}                 
 \def\Pb      {\ensuremath{\mathrm{b}}\xspace}                 
                  
 \def\Pd      {\ensuremath{\mathrm{d}}\xspace}                 
 \def\Pe      {\ensuremath{\mathrm{e}}\xspace}                 
 \def\Ps      {\ensuremath{\mathrm{s}}\xspace}                 
 \def\Pt      {\ensuremath{\mathrm{t}}\xspace}                 

 \def\thebaroffset{0.0em}
}
{

 \def\Ppi         {\ensuremath{\pi}\xspace}                 
                  
 \def\Prho        {\ensuremath{\rho}\xspace}

 \mathchardef\PDelta="7101
 \mathchardef\PXi="7104
 \mathchardef\PLambda="7103
 \mathchardef\PSigma="7106
 \mathchardef\POmega="710A
 \mathchardef\PUpsilon="7107
 \mathchardef\PPi="7105
 \def\PB      {\ensuremath{B}\xspace}                 
 \def\PD      {\ensuremath{D}\xspace}                 

 \def\PK      {\ensuremath{K}\xspace}                 
 \def\Pb      {\ensuremath{b}\xspace}                 
                  
 \def\Pd      {\ensuremath{d}\xspace}                 
 \def\Pe      {\ensuremath{e}\xspace}                 
 \def\Ps      {\ensuremath{s}\xspace}                 
 \def\Pt      {\ensuremath{t}\xspace}                 

 \def\thebaroffset{0.18em}
}
\newcommand{\offsetoverline}[2][\thebaroffset]{\kern #1\overline{\kern -#1 #2}}%

\makeatletter
\ifcase \@ptsize \relax% 10pt
  \newcommand{\miniscule}{\@setfontsize\miniscule{4}{5}}% \tiny: 5/6
\or% 11pt
  \newcommand{\miniscule}{\@setfontsize\miniscule{5}{6}}% \tiny: 6/7
\or% 12pt
  \newcommand{\miniscule}{\@setfontsize\miniscule{5}{6}}% \tiny: 6/7
\fi
\makeatother

\DeclareRobustCommand{\optbar}[1]{\shortstack{{\miniscule (\rule[.5ex]{1.25em}{.18mm})}
  \\ [-.7ex] $#1$}}

\def\en         {{\ensuremath{\Pe^-}}\xspace}   % electron negative (\em is taken)
\def\ep         {{\ensuremath{\Pe^+}}\xspace}

\def\dquark    {{\ensuremath{\Pd}}\xspace}

\def\squark    {{\ensuremath{\Ps}}\xspace}

\def\bquark    {{\ensuremath{\Pb}}\xspace}

\def\tquark    {{\ensuremath{\Pt}}\xspace}

\def\pion   {{\ensuremath{\Ppi}}\xspace}
\def\piz    {{\ensuremath{\pion^0}}\xspace}
\def\pip    {{\ensuremath{\pion^+}}\xspace}
\def\pim    {{\ensuremath{\pion^-}}\xspace}

\def\rhomeson {{\ensuremath{\Prho}}\xspace}

\def\rhop     {{\ensuremath{\rhomeson^+}}\xspace}

\def\kaon    {{\ensuremath{\PK}}\xspace}
\def\Kbar    {{\ensuremath{\offsetoverline{\PK}}}\xspace}

\def\KorKbar {\kern \thebaroffset\optbar{\kern -\thebaroffset \PK}{}\xspace}

\def\Kp      {{\ensuremath{\kaon^+}}\xspace}
\def\Km      {{\ensuremath{\kaon^-}}\xspace}

\def\Kstar   {{\ensuremath{\kaon^*}}\xspace}
\def\Kstarb  {{\ensuremath{\Kbar{}^*}}\xspace}

\def\D       {{\ensuremath{\PD}}\xspace}

\def\DorDbar {\kern \thebaroffset\optbar{\kern -\thebaroffset \PD}\xspace}

\def\Dp      {{\ensuremath{\D^+}}\xspace}
\def\Dm      {{\ensuremath{\D^-}}\xspace}

\def\DpDm    {\ensuremath{\Dp {\kern -0.16em \Dm}}\xspace}

\def\B       {{\ensuremath{\PB}}\xspace}
\def\Bbar    {{\ensuremath{\offsetoverline{\PB}}}\xspace}

\def\BorBbar {\kern \thebaroffset\optbar{\kern -\thebaroffset \PB}\xspace}

\def\Bzb     {{\ensuremath{\Bbar{}^0}}\xspace}
\def\Bd      {{\ensuremath{\B^0}}\xspace}

\def\BdorBdbar {\kern \thebaroffset\optbar{\kern -\thebaroffset \Bd}\xspace}
\def\Bu      {{\ensuremath{\B^+}}\xspace}

\def\Bp      {{\ensuremath{\Bu}}\xspace}

\def\Bs      {{\ensuremath{\B^0_\squark}}\xspace}

\def\BsorBsbar {\kern \thebaroffset\optbar{\kern -\thebaroffset \Bs}\xspace}

\def\Bds     {{\ensuremath{\B_{(\squark)}^0}}\xspace}

\def\BdorBs  {\Bds}

\def\Y#1S{\ensuremath{\PUpsilon{(#1S)}}\xspace}

\def\Lz          {{\ensuremath{\PLambda}}\xspace}

\def\LorLbar     {\kern \thebaroffset\optbar{\kern -\thebaroffset \PLambda}\xspace}

\def\Lb           {{\ensuremath{\Lz^0_\bquark}}\xspace}

\def\BF         {{\ensuremath{\mathcal{B}}}\xspace}

\newcommand{\decay}[2]{\ensuremath{\mathinner{#1\!\to #2}}\xspace}

\def\to                 {\ensuremath{\rightarrow}\xspace}

\def\grpsuthree {{\ensuremath{\mathrm{SU}(3)}}\xspace}

\def\Vtd  {{\ensuremath{V_{\tquark\dquark}^{\phantom{\ast}}}}\xspace}

\def\Vts  {{\ensuremath{V_{\tquark\squark}^{\phantom{\ast}}}}\xspace}

\def\Vtds  {{\ensuremath{V_{\tquark\dquark}^\ast}}\xspace}

\def\Vtss  {{\ensuremath{V_{\tquark\squark}^\ast}}\xspace}

\def\AT#1     {\ensuremath{A_{\mathrm{T}}^{#1}}\xspace}           % 2

\def\C#1      {\ensuremath{\mathcal{C}_{#1}}\xspace}                       % 9
\def\Cp#1     {\ensuremath{\mathcal{C}_{#1}^{'}}\xspace}                    % 7
\def\Ceff#1   {\ensuremath{\mathcal{C}_{#1}^{\mathrm{(eff)}}}\xspace}        % 9  
\def\Cpeff#1  {\ensuremath{\mathcal{C}_{#1}^{'\mathrm{(eff)}}}\xspace}       % 7
\def\Ope#1    {\ensuremath{\mathcal{O}_{#1}}\xspace}                       % 2
\def\Opep#1   {\ensuremath{\mathcal{O}_{#1}^{'}}\xspace}                    % 7

\newcommand{\aunit}[1]{\ensuremath{\text{\,#1}}}       
\newcommand{\tev}{\aunit{Te\kern -0.1em V}\xspace}
\newcommand{\gev}{\aunit{Ge\kern -0.1em V}\xspace}
\newcommand{\mev}{\aunit{Me\kern -0.1em V}\xspace}
\newcommand{\kev}{\aunit{ke\kern -0.1em V}\xspace}
\newcommand{\ev}{\aunit{e\kern -0.1em V}\xspace}
 
\newcommand{\mevc}{\ensuremath{\aunit{Me\kern -0.1em V\!/}c}\xspace}
\newcommand{\gevc}{\ensuremath{\aunit{Ge\kern -0.1em V\!/}c}\xspace}
\newcommand{\mevcc}{\ensuremath{\aunit{Me\kern -0.1em V\!/}c^2}\xspace}
\newcommand{\gevcc}{\ensuremath{\aunit{Ge\kern -0.1em V\!/}c^2}\xspace}
\def\fb   {\ensuremath{\aunit{fb}}\xspace}
\def\invfb   {\ensuremath{\fb^{-1}}\xspace}

\def\gsim{{~\raise.15em\hbox{$>$}\kern-.85em
          \lower.35em\hbox{$\sim$}~}\xspace}
\def\lsim{{~\raise.15em\hbox{$<$}\kern-.85em
          \lower.35em\hbox{$\sim$}~}\xspace}

\def\pt         {\ensuremath{p_{\mathrm{T}}}\xspace}

\def\et         {\ensuremath{E_{\mathrm{T}}}\xspace}

\def\evtgen     {\mbox{\textsc{EvtGen}}\xspace}

\def\geant      {\mbox{\textsc{Geant4}}\xspace}

\def\photos     {\mbox{\textsc{Photos}}\xspace}

\def\pythia     {\mbox{\textsc{Pythia}}\xspace}

\def\tell1  {TELL1\xspace}
\def\ukl1   {UKL1\xspace}

\newcommand{\lhcborcid}[1]{\href{https://orcid.org/#1}{\hspace*{0.1em}\raisebox{-0.45ex}{\includegraphics[width=1em]{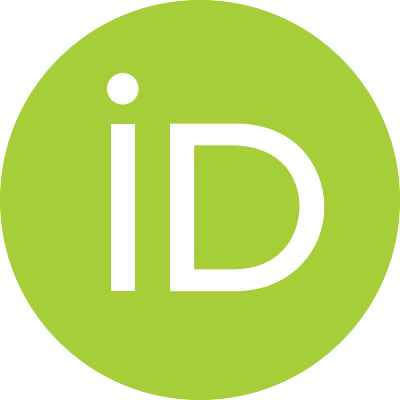}}}}

\hypersetup{
  colorlinks   = true, %Colours links instead of ugly boxes
  urlcolor     = blue, %Colour for external hyperlinks
  linkcolor    = blue, %Colour of internal links
  citecolor    = red   %Colour of citations
}

\ifEnableSectionTOCLinks
    \usepackage[explicit]{titlesec} % to change headings
    
    \let\oldcontentsline\contentsline
    \renewcommand

    \titleformat{\section}{\normalfont\Large\bf}{\hyperlink{tocsection.\thesection}{{\thesection} \parbox[t]{\dimexpr\textwidth-1pc}{#1}}}{1pc}{}

    \titleformat{\subsection}{\normalfont\bf}{\hyperlink{tocsubsection.\thesubsection}{{\thesubsection} \parbox[t]{\dimexpr\textwidth-1pc}{#1}}}{1pc}{}

    \titleformat{name=\section,numberless}[display]{}{}{0pt}{\normalfont\Huge\bfseries #1}
\fi

\usepackage{cite} % Allows for ranges in citations
\usepackage{LHCb/mciteplus}
\usepackage{multirow}
\usepackage{subcaption}

\def\kpi{{\ensuremath{\Km\pip}}\xspace}
\def\mkpi{{\mbox{\ensuremath{m(\kpi)}}}\xspace}
\def\mkpilow{{\mbox{\ensuremath{796<m(\kpi)<996\mevcc}}}\xspace}
\def\mkpihigh{{\mbox{\ensuremath{996<m(\kpi)<1800\mevcc}}}\xspace}

\def\bdkpig{{\mbox{\ensuremath{\decay{\Bzb}{\Km\pip\gamma}}}}\xspace}
\def\bskpig{{\mbox{\ensuremath{\decay{\Bs}{\Km\pip\gamma}}}}\xspace}
\def\bdskpig{{\mbox{\ensuremath{\decay{\BdorBsorbar}{\Km\pip\gamma}}}}\xspace}
\def\bskkg{{\mbox{\ensuremath{\decay{\Bs}{\Kp\Km\gamma}}}}\xspace}
\def\lbpkg{{\mbox{\ensuremath{\decay{\Lb}{p\Km\gamma}}}}\xspace}
\def\kpig{{\mbox{\ensuremath{\kpi\gamma}}}\xspace}
\def\kkg{{\mbox{\ensuremath{\Kp\Km\gamma}}}\xspace}
\def\pkg{{\mbox{\ensuremath{p\Km\gamma}}}\xspace}
\def\fsfd{{\mbox{\ensuremath{\frac{f_s}{f_d}}}}\xspace}

\def\BdorBsorbar {\kern-0.4em\optbar{\kern0.4em \BdorBs}\xspace}
\def\Kstarzv{{\mbox{\ensuremath{\Kstar(892)^0}}}\xspace}
\def\Kstarzbv{{\mbox{\ensuremath{\Kstarb(892)^0}}}\xspace}
\def\bskstg{{\mbox{\ensuremath{\decay{\Bs}{\Kstarzbv\gamma}}}}\xspace}

\begin{document}

%%%%%%%%%%%%%%%%%%%%%%%%%
%%%%% Title     %%%%%%%%%
%%%%%%%%%%%%%%%%%%%%%%%%%
\renewcommand{\thefootnote}{\fnsymbol{footnote}}
\setcounter{footnote}{1}

% ===============================================================================
% Purpose: LHCb-PAPER journal paper title page template
% Author: 
% Created on: 2010-09-25
% ===============================================================================

%%%%%%%%%%%%%%%%%%%%%%%%%
%%%%%  TITLE PAGE  %%%%%%
%%%%%%%%%%%%%%%%%%%%%%%%%
\begin{titlepage}
\pagenumbering{roman}

% Header ---------------------------------------------------
\vspace*{-1.5cm}
\centerline{\large EUROPEAN ORGANIZATION FOR NUCLEAR RESEARCH (CERN)}
\vspace*{1.5cm}
\noindent
\begin{tabular*}{\linewidth}{lc@{\extracolsep{\fill}}r@{\extracolsep{0pt}}}
\ifthenelse{\boolean{pdflatex}}% Logo format choice
{\vspace*{-1.5cm}\mbox{\!\!\!\includegraphics[width=.14\textwidth]{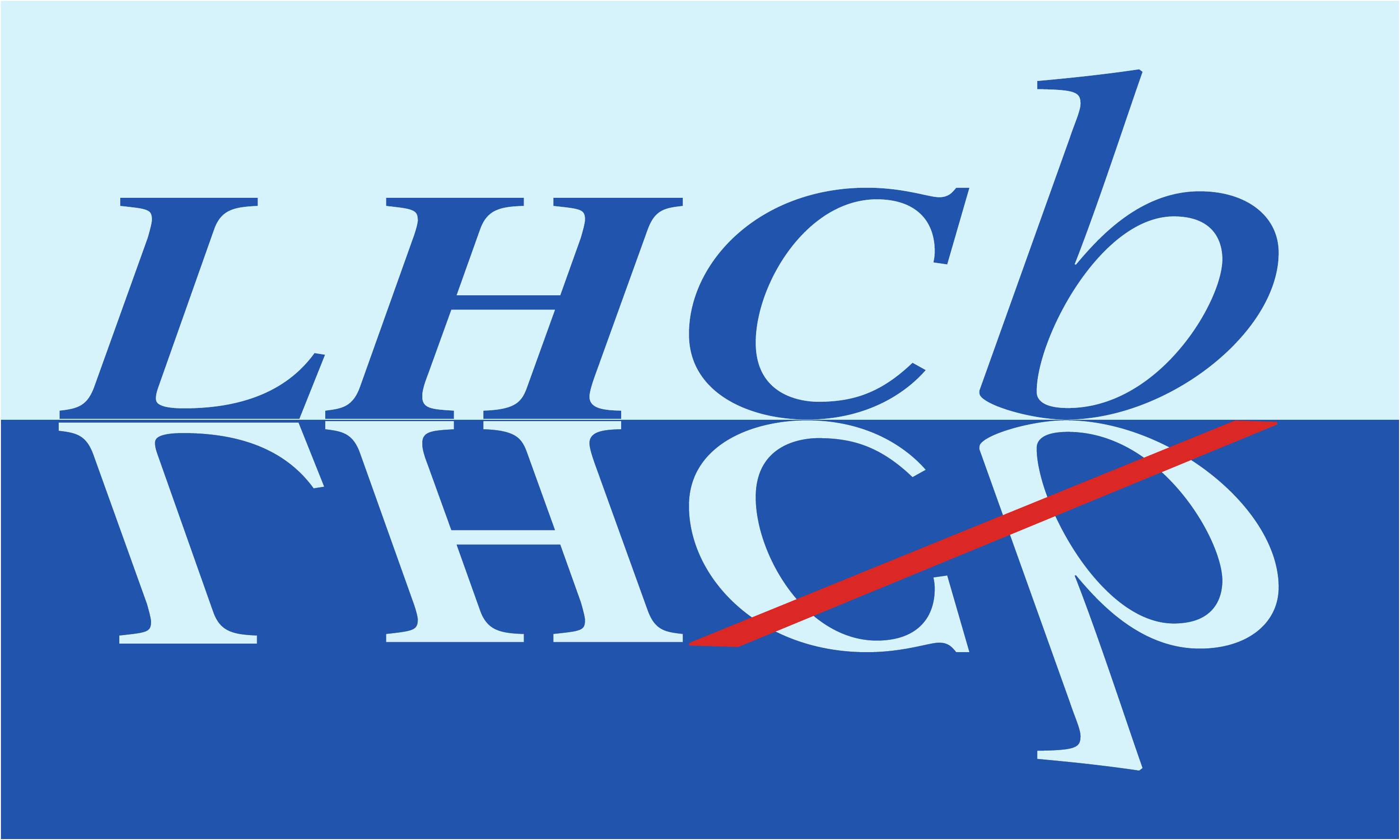}} & &}%
{\vspace*{-1.2cm}\mbox{\!\!\!\includegraphics[width=.12\textwidth]{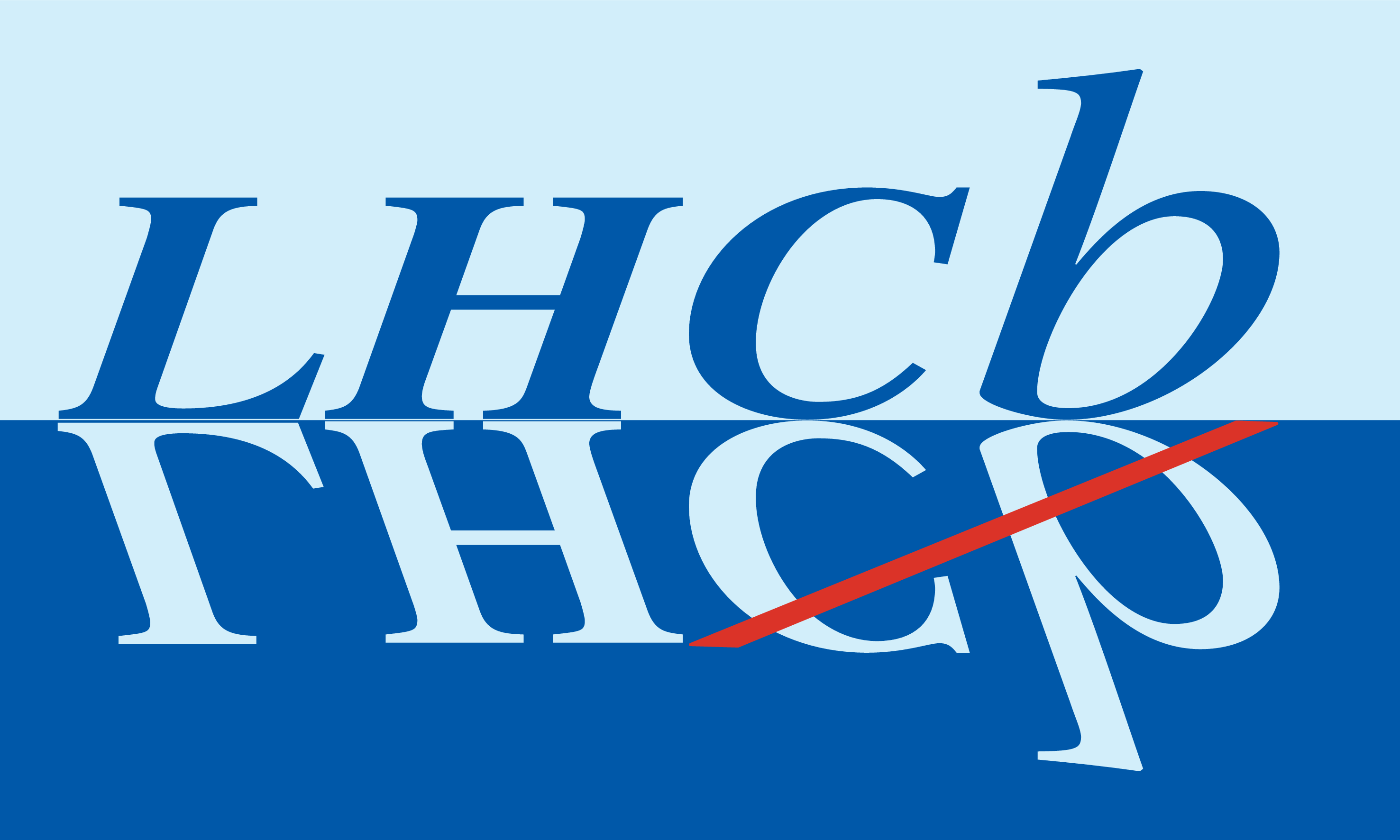}} & &}%
\\
 & & CERN-EP-2025-276 \\  % ID 
 & & LHCb-PAPER-2025-056 \\  % ID 
 & & April 30, 2026 \\ % Date - Can also hardwire e.g.: 23 March 2010
% not in paper \hline
\end{tabular*}

\vspace*{4.0cm}

% Title --------------------------------------------------
{\normalfont\bfseries\boldmath\huge
\begin{center}
% DO NOT EDIT HERE. Instead edit macro in main.tex to keep metadata correct
  \papertitle 
\end{center}
}

\vspace*{2.0cm}

% Authors -------------------------------------------------
\begin{center}
% Edit macro in main.tex to keep metadata correct
\paperauthors\footnote{Authors are listed at the end of this Letter.}
\end{center}

\vspace{\fill}

% Abstract -----------------------------------------------
\begin{abstract}
  \noindent
The first search for the $B_s^0\rightarrow K^-\pi^+\gamma$ decay in the range $796<m(K^-\pi^+)<1800\,\text{Me\kern -0.1em V\!/}c^2$ is performed using data from proton-proton collisions collected by the LHCb experiment at centre-of-mass energies of 7, 8, and 13~TeV, corresponding to an integrated luminosity
of 9~fb$^{-1}$.
The photons are reconstructed through their conversion into an electron-positron pair, which significantly improves the mass resolution of the reconstructed decays with respect to decays with an unconverted photon.
A signal excess with a significance of 3.5 standard deviations is measured, constituting the first experimental evidence for this decay.
In the range $796<m(K^-\pi^+)<996\,\text{Me\kern -0.1em V\!/}c^2$, the ratio $\mathcal R$ between the branching fractions of the signal decay and the favoured $\kern 0.18em\overline{\kern -0.18em B}{}^0\rightarrow K^- \pi^+\gamma$ decay is measured to be \mbox{${\mathcal R} = (3.7\pm1.2\pm0.4)\times10^{-2}$} where the first uncertainty is statistical and the second is systematic.
This measurement is consistent with the value predicted in the Standard Model.
In the range $996<m(K^-\pi^+)<1800\,\text{Me\kern -0.1em V\!/}c^2$, the ratio \mbox{${\mathcal R} = (0.2\pm2.7\pm1.3)\times10^{-2}$} is measured.
  
\end{abstract}

\vspace*{2.0cm}

\begin{center}
  Published in
  Phys.~Rev.~Lett.
\end{center}

\vspace{\fill}

{\footnotesize 
% Edit macro in main.tex to keep metadata correct
\centerline{\copyright~\papercopyright. \href{\paperlicenceurl}{\paperlicence}.}}
\vspace*{2mm}

\end{titlepage}

%%%%%%%%%%%%%%%%%%%%%%%%%%%%%%%%
%%%%%  EOD OF TITLE PAGE  %%%%%%
%%%%%%%%%%%%%%%%%%%%%%%%%%%%%%%%

%  empty page follows the title page ----
\newpage
\setcounter{page}{2}
\mbox{~}

%\twocolumn
% %%%%%%%%%%%%% ---------

\renewcommand{\thefootnote}{\arabic{footnote}}
\setcounter{footnote}{0}

%%%%%%%%%%%%%%%%%%%%%%%%%%%%%%%%
%%%%%  Table of Content   %%%%%%
%%%%%%%%%%%%%%%%%%%%%%%%%%%%%%%%

\cleardoublepage

%%%%%%%%%%%%%%%%%%%%%%%%%
%%%%% Main text %%%%%%%%%
%%%%%%%%%%%%%%%%%%%%%%%%%

\pagestyle{plain} % restore page numbers for the main text
\setcounter{page}{1}
\pagenumbering{arabic}

%% This is the main body
%% It is useful to have a single file so comments are not missed in overleaf.
In the Standard Model (SM), \decay{\bquark}{\dquark\gamma} and \decay{\bquark}{\squark\gamma} quark transitions proceed via electroweak loop diagrams.\footnote{The inclusion of charge-conjugate processes is implied throughout this Letter.}
Particles beyond the SM (BSM) of arbitrarily high mass can contribute to these diagrams~\cite{Li:2013vlx,Xiao:2003vq,Ali:2002kw,Arhrib:2001gr}.
The \decay{\bquark}{\squark\gamma} transition, with amplitude proportional to the Cabibbo--Kobayashi--Maskawa (CKM) matrix element \Vtss,  has been measured through several radiative \bquark-hadron decays~\cite{LHCb-PAPER-2019-015, LHCb-PAPER-2021-030, LHCb-PAPER-2020-020, LHCb-PAPER-2024-030, LHCb-PAPER-2012-019, Belle-II:2024kbc, BaBar:2015chw, Paul:2016urs}.
In contrast, measurements of the \decay{\bquark}{\dquark\gamma} transition, with amplitude proportional to \Vtds, remain limited.
Branching fractions (\BF) of the \decay{\B^{+/0}}{\rho(770)^{+/0}\gamma} and \decay{\Bd}{\omega(782)\gamma} decays have been measured~\cite{LHCb-PAPER-2025-017,Belle:2024qhe,Belle:2008imr,BaBar:2008txc} as well as the inclusive branching fraction of \mbox{\decay{B}{X_{d}\gamma}} decays~\cite{BaBar:2010vgu}, where $X_d$ is an unflavoured hadronic system.

The ratio of branching fractions between decays mediated by \decay{\bquark}{\dquark\gamma} and \decay{\bquark}{\squark\gamma} transitions can be used to measure $|\Vtd/\Vts|$~\cite{Ball:2006nr,Ball:2006eu}.
This value may be compared to determinations based on \BdorBs meson oscillations~\cite{LHCb-PAPER-2021-005,HFLAV23} to identify potential contributions from BSM processes.
The value of $|\Vtd/\Vts|$ extracted from the measured branching fraction ratio \mbox{$\mathcal{R}\equiv\BF(\decay{\Bs}{\Kstarzbv\gamma})/\BF(\decay{\Bzb}{\Kstarzbv\gamma})$} is theoretically cleaner than determinations using other decays involving \decay{\bquark}{\dquark\gamma} transitions since dominant uncertainties from the ratios of form factors largely cancel~\cite{Ball:2006eu}.
Using the existing calculations of the form-factor ratio and nonfactorisable effects~\cite{Ball:2006eu} with updated meson lifetimes, masses and CKM matrix elements~\cite{CKMfitter2005,CKMfitter2015,PDG2024}, the prediction is \mbox{$\mathcal{R}=(3.9\pm0.7)\times10^{-2}$}.
However, a new calculation of the form-factor ratio and nonfactorisable effects, incorporating the latest experimental information, will be necessary to reduce the uncertainty on an eventual measurement of $|\Vtd/\Vts|$ using \bskstg decays.

This Letter reports the first search and first evidence for the \bskpig decay.
The reconstructed $\Km\pip$ system includes contributions from the \Kstarzbv resonance as well as higher-mass states up to 1800\mevcc.
Additionally, the ratio of its branching fraction with respect to the \bdkpig decay is measured.
The analysis uses proton-proton ($pp$) collision data recorded by the LHCb experiment, corresponding to integrated luminosities of $3\invfb$ at centre-of-mass energies of 7 and 8\tev, and $6\invfb$ at 13\tev, referred to as the {Run~1} and {Run~2} data samples, respectively.
In order to avoid experimenter's bias, the data in a mass window within $\pm 40\mevcc$ of the known mass of the \Bs meson~\cite{PDG2024}, corresponding to approximately twice the resolution of the signal decay, were not examined until the full procedure had been finalised.

The LHCb detector~\cite{LHCb-DP-2008-001,LHCb-DP-2014-002} is a single-arm forward spectrometer covering the pseudorapidity range $2 < \eta < 5$.
Detector elements include: a silicon-strip vertex detector (VELO) surrounding the $pp$ interaction region~\cite{LHCb-DP-2014-001}; a tracking system, consisting of detectors upstream and downstream of a dipole magnet with a bending power of about $4{\mathrm{\,T\,m}}$, that provides a measurement of the momentum, $p$, of charged particles~\cite{LHCb-DP-2013-003,LHCb-DP-2017-001}; two ring-imaging Cherenkov (RICH) detectors that are able to discriminate between different species of charged hadrons~\cite{LHCb-DP-2012-003}; and a calorimeter system able to identify and measure the energy of electrons and photons~\cite{LHCb-DP-2020-001}.

The online event selection is performed by a trigger~\cite{LHCb-DP-2012-004,LHCb-DP-2019-001}.
At the hardware trigger stage, events are required to have a muon with high transverse momentum (\pt) or a hadron, photon or electron with high transverse energy (\et) in the calorimeters.
The software trigger performs a full event reconstruction and requires that the signal decay candidates contain a secondary vertex consistent with the decay of a \bquark hadron formed of either multiple tracks~\cite{BBDT,LHCb-PROC-2015-018} or two high-\pt tracks alongside a high-\et photon~\cite{LHCb-PUB-2012-002}.

Simulation is required to model the effects of the detector resolution and acceptance, and the imposed selection requirements.
The $pp$ collisions are generated using \pythia~\cite{Sjostrand:2007gs,*Sjostrand:2006za} with a specific \lhcb configuration~\cite{LHCb-PROC-2010-056}.
Decays of unstable particles are described by \evtgen~\cite{Lange:2001uf}, in which final-state radiation is generated using \photos~\cite{davidson2015photos}.
The interaction of the generated particles with the detector, and its response, are implemented using the \geant toolkit~\cite{Allison:2006ve, *Agostinelli:2002hh} as described in Ref.~\cite{LHCb-PROC-2011-006}. 
The simulated samples of \Bd, \Bs and \Lb decays are corrected to account for data-simulation differences in the \bquark-hadron production kinematics and detector occupancy using data samples enriched in \bdkpig, \bskkg, and \lbpkg decays, respectively.
Particle-identification efficiencies are calibrated using dedicated control samples~\cite{LHCb-DP-2018-001}.

Only photons that convert through interactions with the detector material into an electron-positron pair, referred to as \emph{converted photons}~\cite{LHCb-PAPER-2013-028,LHCb-PAPER-2014-040,LHCb-PAPER-2015-044}, are used in this search.
This improves the mass resolution of the \bquark-hadron candidate by a factor of three with respect to candidates formed using photons reconstructed as clusters in the electromagnetic calorimeter (ECAL), reducing the background from the \bdkpig decay. Furthermore, the photon trajectory can be matched to the position of the $\Km\pip$ combination vertex to reduce background from random combinations of photons and hadrons.
Electron candidates are identified among high-\pt tracks that are associated with clusters in the ECAL.
To mitigate loss in resolution from bremsstrahlung before the magnet, electrons are rejected if an ECAL cluster, identifiable as a photon with $\et>75\mev$, is found at a position consistent with the extrapolation of the electron trajectory.
Converted photons are reconstructed in two categories: those that convert early enough that the electrons are reconstructed using the full tracking system (\emph{long} tracks); and those that convert further downstream, where no track segments are formed in the VELO (\emph{downstream} tracks).
To reject background from random combinations, the two electron candidates are required to originate from a common vertex, and the mass of the dielectron must be less than \mbox{50\mevcc}.
The photon candidate must have $\pt>500\mevc$ and its vertex must be significantly displaced from the primary vertex~(PV) that is most consistent with the flight direction of the \B candidate, referred to as the associated PV.

To evaluate the contamination to the \bdskpig signal decays due to particle misidentification, \bskkg and \lbpkg decays are also analysed.
All of these decay candidates are constructed by combining two high-\pt tracks incompatible with originating from any PV and identified using information from the RICH system as pions, kaons or protons, and requiring that they form a common vertex.
The photon candidate is combined with the dihadron pair consistent with originating from the same vertex to form decay candidates that are consistent with originating from a PV.
The dihadron masses are required to be in the ranges \mbox{$796<m(\kpi)<1800\mevcc$},  \mbox{$1000<m(\Kp\Km)<1800\mevcc$} and \mbox{$1400<m(p\Km)<2500\mevcc$}.
To reject backgrounds from decays involving a \piz meson misidentified as a photon, such as the \decay{\Bzb}{\left[\pip\piz\right]_{\rhop}\Km} decay, the mass of the combination of the photon assigned the \piz mass hypothesis with one hadron is required to be greater than $2000\mevcc$.
Momentum resolution is improved by applying a kinematic fit~\cite{Hulsbergen:2005pu}, in which all final-state particles are constrained to originate from the corresponding decay vertex, and the \mbox{$b$-hadron} candidate is constrained to originate from the associated PV.
The mass resolution is further improved by requiring the mass of downstream photon candidates to be below $40\mevcc$ and, for candidates containing long electron tracks, the uncertainty on the \bquark-hadron candidate mass to be less than $20\mevcc$. These requirements are approximately 85\% and 90\% efficient on signal decays, respectively.

The data are analysed separately for the Run~1 and 2 datasets, long and downstream electrons, and low (\mkpilow) and high (\mkpihigh) dihadron mass windows.
Separate boosted decision tree~(BDT) classifiers~\cite{Breiman,AdaBoost,Hocker:2007ht,*TMVA4} are trained for each of these categories to further reject background from random combinations of particles.
The classifiers are trained using simulated \mbox{\decay{\Bs}{\Kstarzbv\gamma}} decays for the signal and \bdskpig candidates in data with masses above $5580\mevcc$ for the background sample. 
The classifiers exploit kinematic and topological properties of the decay as well as, for long electron tracks, the amount of charge deposited in the VELO that is larger for converted photons where the two electron tracks overlap, and their compatibility with the electron hypothesis using information from the RICH and calorimeter systems.
The requirement on the classifier in the \kpig final state is optimised simultaneously with requirements on the kaon and pion-identification probabilities to maximise the figure of merit \mbox{$\varepsilon_{\rm sig}/(3/2 + \sqrt{N_{\rm bkg}})$}~\cite{Punzi:2003bu} for the purpose of finding evidence for this decay.
Here, $\varepsilon_{\rm sig}$ denotes the efficiency of the selection requirements determined from simulation, and $N_{\rm bkg}$ is the estimated number of candidates from all processes except the signal decay within a mass window corresponding to four times the signal mass resolution, centred on its peak position.
The optimal particle identification and BDT requirements correspond to a combinatorial background rejection of at least 95\% and a signal efficiency of 55--98\%, where less efficient requirements are necessary for data in \mkpihigh where signal purity is lower.
The same BDT requirement is applied to the \kkg and \pkg final states without any additional particle-identification criteria.

\begin{figure}[tb]
  \centering
      \includegraphics[width=\linewidth]{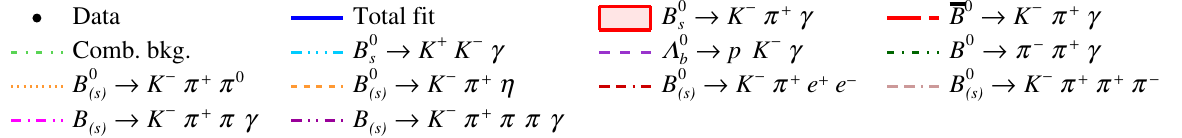}\\
  \includegraphics[width=\linewidth/2]{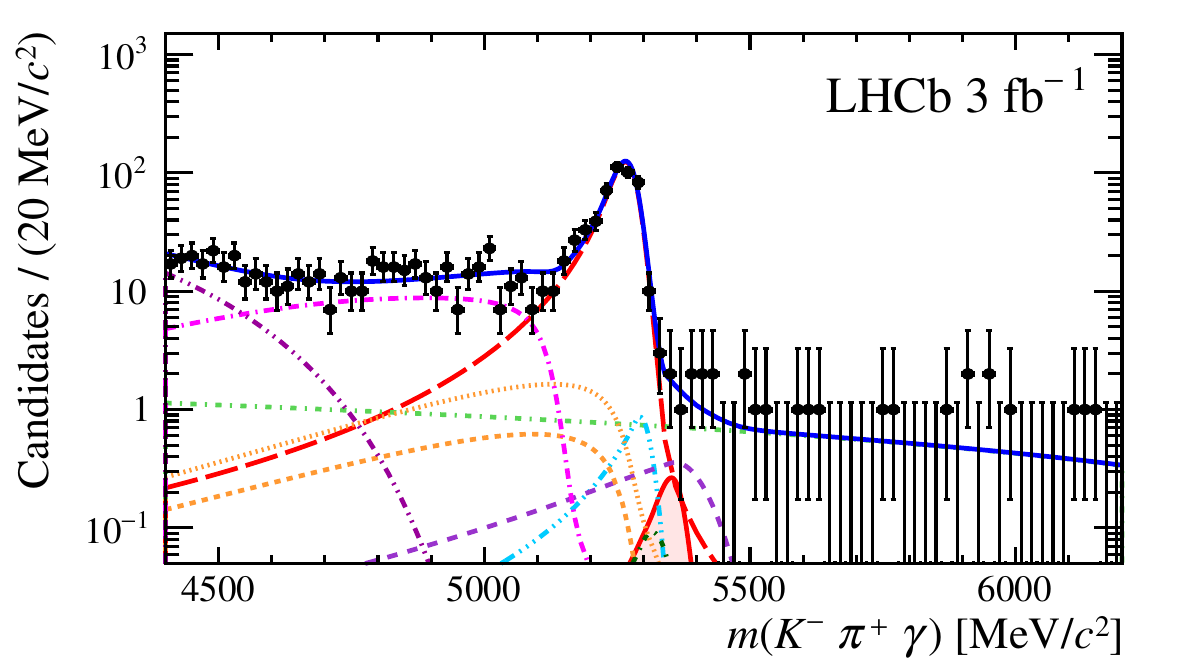}%
  \includegraphics[width=\linewidth/2]{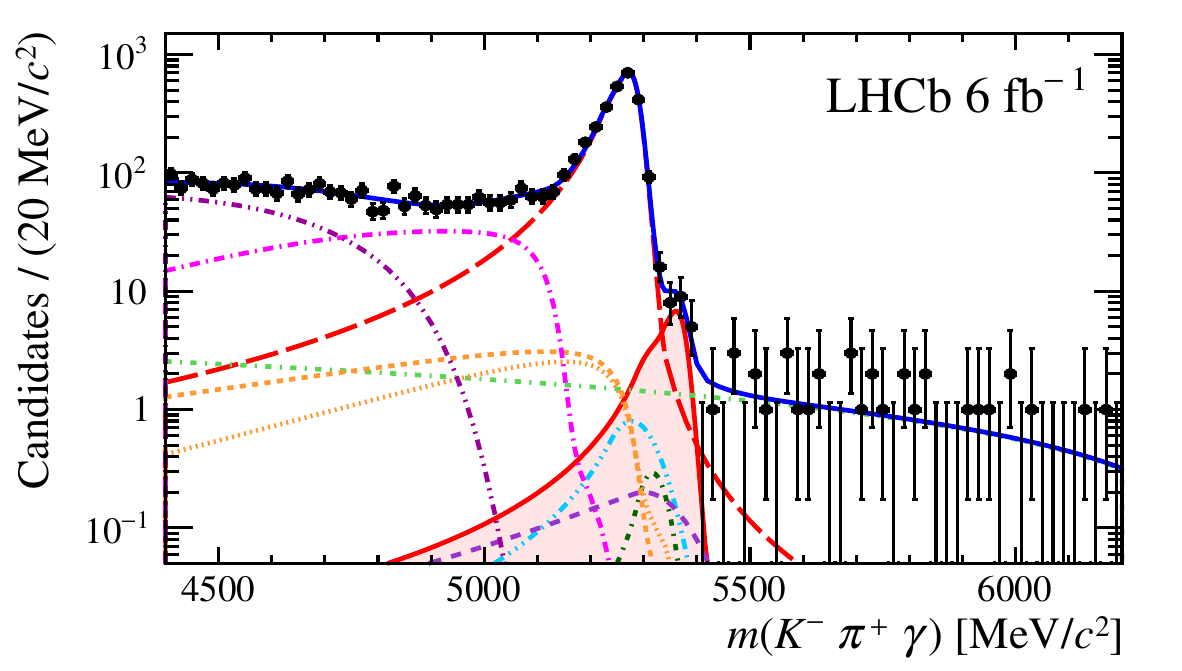}\\
  \includegraphics[width=\linewidth/2]{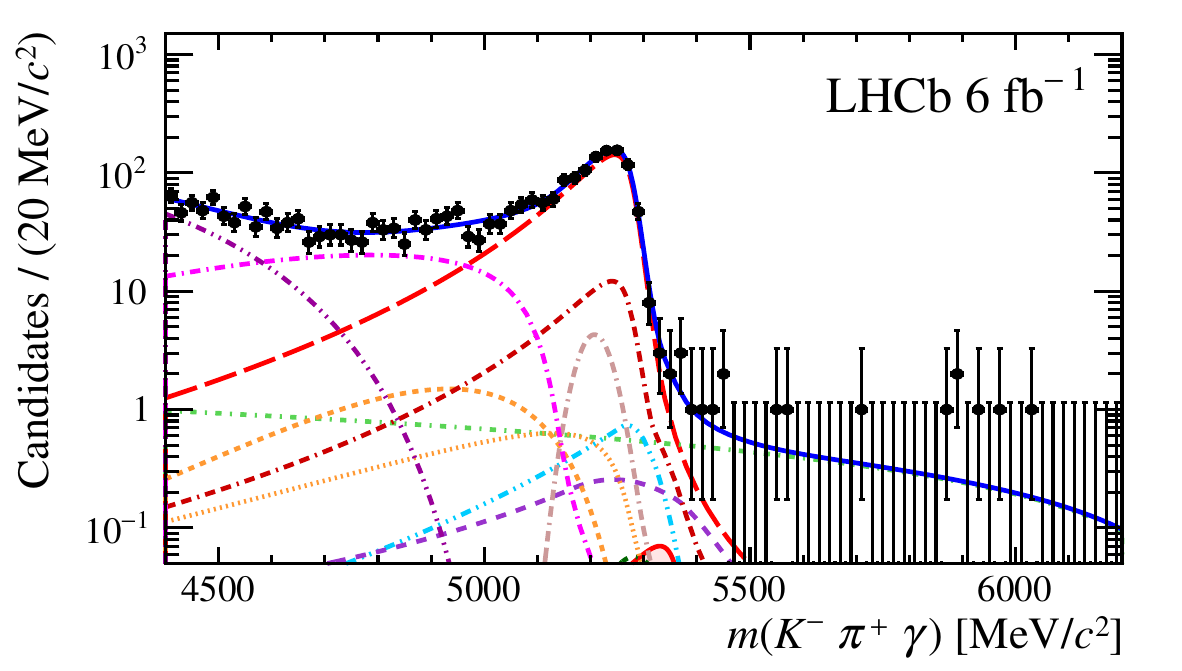}%
  \includegraphics[width=\linewidth/2]{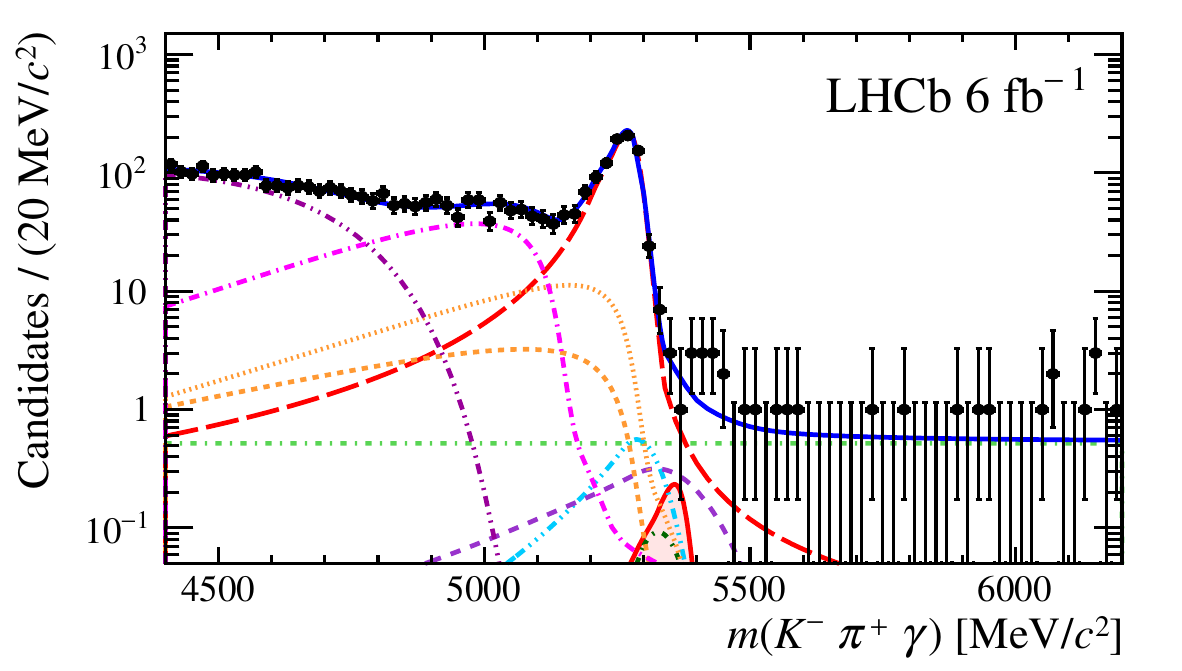}
  \caption{\small Mass distribution of selected $K^-\pi^+\gamma$ candidates in (top left) Run~1 downstream low-$m(K^-\pi^+)$ (\mkpilow); (top right) Run~2 downstream low-$m(K^-\pi^+)$; (bottom left) Run~2 long low-$m(K^-\pi^+)$ and (bottom right) Run~2 downstream high-$m(K^-\pi^+)$ (\mkpihigh) datasets.
  The model described in the text, fitted separately to each dataset, is also shown. The corresponding measured $B_s^0 \rightarrow K^- \pi^+ \gamma$ yields are $1\pm5$, $32\pm11$, $2\pm12$ and $0\pm6$, and the measured  $\kern 0.18em\overline{\kern -0.18em B}{}^0\rightarrow K^- \pi^+\gamma$ yields are $523\pm26$, $3\,132\pm61$, $1\,247\pm45$, and $988\pm35$, where the uncertainties are statistical only.}
  \label{fig:datafit}
\end{figure}

The best sensitivity to the signal decay is obtained using downstream electrons, which have a smaller probability of bremsstrahlung and therefore better momentum resolution, and in the low-mass window occupied by the \Kstarzv resonance that offers higher purity. 
Of the eight available subsets of data, only those with the highest signal-to-background purity in data and mass resolution of the normalisation decay are analysed: low-\mkpi with downstream electrons in Run~1 data; both low- and high-\mkpi with downstream electrons in Run~2 data; and low-\mkpi with long track electrons in Run~2 data. The chosen datasets, established prior to examination of the signal region, each provide anexpected signal significance of approximately 1.0 standard deviations or higher assuming the predicted branching fraction; the excluded datasets provide expected significances below 0.3 standard deviations.
Figure~\ref{fig:datafit} shows the mass distributions of \kpig candidates passing the selection requirements in the datasets with best sensitivity.
The signal decay yields are extracted through an extended unbinned maximum-likelihood fit to these data samples.

The \bdskpig signal decays are modelled using Johnson $S_U$ (JSU) functions featuring power-law tails on both sides of the peak~\cite{JohnsonSU}.
For candidates where the downstream tracks are not matched to the correct hits in the tracking detectors, the mass distribution is shifted towards lower values.
To account for this effect, a modified Gaussian function with power-law tails on both sides (double-sided Crystal Ball, DSCB)~\cite{Skwarnicki:1986xj} is added to the JSU function.
All shape parameters are fixed from simulation, except for the relative fraction of the two components and deviations of the overall width and peak position from simulation. These are unconstrained and shared between the \Bd and \bskpig decay models.

Partially reconstructed \mbox{\decay{\B}{\Km\pip\pi(\pi)\gamma}} decays are modelled with ARGUS functions~\cite{ARGUS:1990hfq} convolved with the signal model.
The endpoints of the ARGUS functions are fixed to the peak position of the signal model reduced by the mass of the missing pions.
The other parameters are fixed from simulation, except for the model for two missing pions where the power parameter is free to vary in the fit to data.
The yields corresponding to partially reconstructed \Bd and \Bp meson decays are free to vary in the fit to data.
To estimate the yields of partially reconstructed \Bs meson decays, the \mbox{\decay{\B}{\Km\pip\pi(\pi)\gamma}} decay yields are scaled by the ratio of \Bs to \B meson production cross-sections~\cite{LHCb-PAPER-2020-046} and the branching fraction ratio estimated by applying \grpsuthree flavour symmetry while accounting for dominant sources of \grpsuthree breaking.

The \mbox{\decay{\BdorBsorbar}{\kpi\piz}} and \mbox{\decay{\BdorBsorbar}{\kpi\eta}} decays, where one of the two photons from the \piz or $\eta$ decay is not reconstructed, are similarly modelled using ARGUS functions convolved with the signal model.
The ARGUS endpoint is set to the peak position of the signal model and the shape parameters are fixed from simulation.
The yields of such components are fixed relative to that of the \bdkpig decay using selection efficiencies from simulation and measured branching fractions, or predictions where measurements are not available~\cite{BaBar:2011vfx,Zhou:2023lbc,Wang:2020saq,PDG2024}.
The yields of \mbox{\decay{\Bs}{K^{*0}_J\eta}} decays \mbox{($J=0,2$)}, where neither measurements nor predictions are available, are estimated similarly to the yield of \mbox{\decay{\Bs}{\Km\pip\pi(\pi)\gamma}} decays.

The \bskkg, \lbpkg and \mbox{\decay{\Bd}{\pip\pim\gamma}} decays, where one or both hadrons are misidentified, are modelled using DSCB functions with shape parameters fixed from simulation.
The yield of \mbox{\decay{\Bd}{\pip\pim\gamma}} decays is fixed relative to the yield of \bdkpig decays using efficiencies from simulation and the known \mbox{\decay{\Bd}{\rho(770)^0\gamma}} branching fraction~\cite{LHCb-PAPER-2025-017}, neglecting contributions from higher-mass resonances.
The yields of the other misidentified decays are determined using data selected in the \kkg and \pkg final states, as described in the End Matter.

Finally, \mbox{\decay{\BdorBsorbar}{\Kstarzbv\ep\en}} and misidentified \mbox{\decay{\BdorBsorbar}{\Kstarzbv\pip\pim}} decays are modelled using a JSU and a Gaussian function, respectively.
The \Bzb decay yields are fixed relative to that of the \bdkpig decay using efficiencies from simulation and measured branching fractions~\cite{Belle:2009zue,BaBar:2008jdv,BaBar:2007nlh,PDG2024}; the yields of the corresponding \Bs decays are further scaled using the same method as for \mbox{\decay{\Bs}{\Km\pip\pi(\pi)\gamma}} decays.
The contamination from \mbox{\decay{\BdorBsorbar}{\Kstarzbv\ep\en}} and \mbox{\decay{\BdorBsorbar}{\Kstarzbv\pip\pim}} decays are negligible for downstream electrons.
Background from random combinations of particles is modelled using a first-order polynomial function with free slope and yield.

\begin{table}[tb]
  \centering
  \caption{Sources of absolute systematic uncertainty on the signal yield. Dashes indicate contributions that are not relevant for that dataset.}
  \label{tab:yieldsys}
  \renewcommand{\arraystretch}{1.2}
  \begin{tabular}{lcccc}
    \hline
    Period & Run~1 & Run~2 & Run~2 & Run~2 \\
    Track & downstream & downstream & long & downstream \\
    \mkpi & low & low & low & high \\
    \hline
    Misidentified-hadron yield & $0.4$  & $0.5$  & $1.1$  & $1.3$ \\
    \kpig and $\Kstarzbv\ep\en$ model & $0.4$  & $1.7$  & $3.0$  & $1.8$ \\
    \decay{\B}{\Km\pip\pi\gamma} resonant structure & $<0.1\mathrel{\phantom{<}}$  & $0.1$  & $0.5$  & $0.2$ \\
    Misidentified-hadron model & $0.4$  & $0.2$  & $0.5$  & $0.9$ \\
    $\Kstarzbv\pip\pim$ model & $-$ & $-$ & $<0.1\mathrel{\phantom{<}}$  & $-$\\
    Combinatorial model & $0.1$  & $0.7$  & $1.1$  & $0.1$ \\
    External inputs & $0.2$  & $0.1$  & $0.5$  & $0.3$ \\
    Simulation sample size & $0.4$  & $0.6$  & $0.9$  & $0.4$ \\
    Hadron-identification corrections & $1.6$  & $0.4$  & $0.1$  & $0.9$ \\
    Electron-identification corrections & $-$ & $-$ & $0.1$  & $-$\\
    Simulation corrections & $0.1$  & $1.7$  & $1.4$  & $0.2$ \\
    \hline
    Total & $1.9$  & $2.7$  & $3.9$  & $2.6$ \\
    \hline
  \end{tabular}
\end{table}

Systematic uncertainties on the signal yield are summarised in Table~\ref{tab:yieldsys}. 
Systematic uncertainties on the yield of the \bdkpig decay are calculated similarly.
The systematic uncertainty associated with the statistical uncertainty of the \bskkg and \lbpkg decay yields is evaluated as the reduction in the uncertainty of the \bskpig decay yield when those contributions are fixed.
All other uncertainties are evaluated from pseudoexperiments generated with an alternative model, taking as the uncertainty the average absolute difference in yield obtained when fitting the baseline model compared to the alternative model.
The alternative models are as follows: 
For \bdskpig and \mbox{\decay{\BdorBsorbar}{\Kstarzbv\ep\en}} decays, and the resolution model for partially reconstructed decays, a different width parameter is used for each side of the DSCB function and the JSU function is replaced by a DSCB function.
For long-track electrons, the model is evaluated using only reconstructed tracks matched to the correct hits in the tracking detectors.
For candidates with downstream electrons, the JSU function is evaluated using only candidates where the photon converts downstream of the VELO and the tracks are matched to the correct hits in the tracking detectors, while the DSCB function is evaluated with the remaining candidates.
The model for partially reconstructed \mbox{\decay{\B}{\Km\pip\pi\gamma}} decays is evaluated from simulated \mbox{\decay{\Bp}{\Kstar(892)^0 \pip\gamma}} decays.
Contributions from decays with misidentified hadrons and \mbox{\decay{\BdorBsorbar}{\Kstarzbv\pip\pim}} decays are alternatively modelled using bifurcated Gaussian and symmetric DSCB functions, respectively.
The combinatorial background is modelled with an exponential function.
Branching fractions and \bquark-hadron masses are varied randomly according to their uncertainties in each pseudoexperiment.
Where assumptions are made on branching fractions in the absence of predictions or measurements, a 100\% uncertainty is assumed.
The contamination from \mbox{\decay{\Bd}{\rho(770)^0\gamma}} decays for \mkpihigh is doubled, and a 100\% uncertainty is assigned, to account for missing contributions from heavier resonances.
Shape parameters and efficiencies are varied randomly according to their covariance matrices.
The uncertainty on the hadron-misidentification corrections includes the size of the calibration sample as well as possible biases introduced by the background subtraction method and variations of the corrections within the phase-space bins in which they are derived.
Other simulation corrections, including those related to electron identification, are small and therefore removed.

A yield of \mbox{$38\pm18$} \bskpig decays is measured, where the uncertainty includes both statistical and systematic sources, of which \mbox{$32\pm11$} correspond to the Run~2 downstream low-\mkpi dataset.
Pseudoexperiments show that the probability that background processes can produce a mass distribution that is less compatible with the background-only model than the data, including both statistical and systematic sources of uncertainty, is \mbox{$2.8\times10^{-4}$}, corresponding to a significance of 3.5 standard deviations.

The signal branching fraction is measured relative to that of the \bdkpig decay as
\begin{equation*}
\mathcal{R} \equiv \frac{\BF(\bskpig)}{\BF(\bdkpig)} = \frac{N(\bskpig)}{N(\bdkpig)} \frac{\varepsilon(\bdkpig)}{\varepsilon(\bskpig)} / \fsfd,
\end{equation*}
where $N$ are the yields, $\varepsilon$ are the efficiencies, and \fsfd is the ratio of production cross-sections evaluated by weighting the \pt-dependent measurement~\cite{LHCb-PAPER-2020-046} over the simulated \pt distribution of \bdkpig decays.
The ratio of the efficiencies is consistent with unity for all datasets.
A correction of 1--8\% is applied for a bias in the measured branching fraction arising from the fit method. It is estimated as the average fit bias observed in pseudoexperiments generated with $\mathcal{R}$ equal to its predicted value.

Systematic uncertainties applied on the normalisation of $\mathcal{R}$ are shown in Table~\ref{tab:normsys}.
The uncertainty on the fit bias is the quadrature sum of its statistical uncertainty, the difference in bias between two methods of determining the average, and the variation in the bias value when pseudodatasets are generated with $\mathcal{R}$ equal to its measured value.
The uncertainty on the \BdorBs lifetimes~\cite{HFLAV23} and from the simulation sample size are propagated to uncertainties on the efficiency ratio.
The uncertainties on the measurement of \fsfd are propagated to the weighted average value.
Correlations between measurements from each dataset arising from the use of common inputs for \BdorBs lifetimes and \fsfd are considered.

\begin{table}[tb]
  \centering
  \caption{\small Sources of systematic uncertainty on the normalisation of the branching fraction ratio in percent.}
  \label{tab:normsys}
  \begin{tabular}{lcccc}
    \hline
    Period & Run~1 & Run~2 & Run~2 & Run~2 \\
    Track & downstream & downstream & long & downstream \\
    \mkpi & low & low & low & high \\
    \hline
Fit bias & $8.8$  & $2.1$  & $3.9$  & $2.4$ \\
Lifetime& $0.4$  & $0.5$  & $0.4$  & $0.6$ \\
Production cross-sections & $3.2$  & $3.4$  & $3.4$  & $3.4$ \\
    Simulation sample size & $1.6$  & $1.5$  & $2.3$  & $1.5$ \\
    \hline
    Total & $9.5$  & $4.3$  & $5.7$  & $4.5$ \\
    \hline
  \end{tabular}
\end{table}

The measured branching fraction ratio is found to be consistent ($p$-value greater than 0.05) between data corresponding to different \mkpi windows, electron track types, data-taking periods and magnet polarities, and under changes in the selection requirements on the BDT classifier, hadron-identification probabilities, and variables that impact the mass resolution.
A simultaneous fit to all datasets with \mkpilow, with a common value of $\mathcal{R}$, results in the measurement
\begin{equation*}
    \mathcal{R}_{\rm low }=(3.7\pm1.2\pm0.4)\times10^{-2} \quad\text{for}\quad \mkpilow,
\end{equation*}
where the first uncertainty is statistical and the second systematic, which is in excellent agreement with the SM prediction.
Additionally, the ratio
\begin{equation*}
    \mathcal{R}_{\rm high}=(0.2\pm2.7\pm1.3)\times10^{-2} \quad\text{for}\quad \mkpihigh,
\end{equation*}
is measured. Correlations between the two measurements are negligible.

In summary, a search for the rare \bskpig decay is performed using $pp$ collision data corresponding to an integrated luminosity of 9\invfb.
An excess of events is found with a significance of 3.5 standard deviations, which constitutes the first evidence for this decay.
Updated theoretical inputs will permit a test of the SM using the measured branching fraction ratio.
Large improvements in precision are anticipated with the new dataset being collected by the upgraded LHCb detector~\cite{LHCb-DP-2022-002}.

% Comment this in for paper drafts; do not include this in analysis note, conference and figure reports
\vspace{\baselineskip}
\emph{Data availability:} Data associated to the plots in this publication are made available on the CERN Document Server in Ref.~\cite{LHCb-PAPER-2025-056-cds}.
\section*{Acknowledgements}
%
% These Acknowledgements valid from 12/09/25
%
\noindent We express our gratitude to our colleagues in the CERN
accelerator departments for the excellent performance of the LHC. We
thank the technical and administrative staff at the LHCb
institutes.
We acknowledge support from CERN and from the national agencies:
ARC (Australia);
CAPES, CNPq, FAPERJ and FINEP (Brazil); 
MOST and NSFC (China); 
CNRS/IN2P3 (France); 
BMFTR, DFG and MPG (Germany);
INFN (Italy); 
NWO (Netherlands); 
MNiSW and NCN (Poland); 
MCID/IFA (Romania); 
%MSHE (Russia); 
MICIU and AEI (Spain);
SNSF and SER (Switzerland); 
NASU (Ukraine); 
STFC (United Kingdom); 
DOE NP and NSF (USA).
%%%%%%%%%%%%%%%%%%%%%%%%%%%%%%%%%%%%%%%%%%%%%
We acknowledge the computing resources that are provided by ARDC (Australia), 
CBPF (Brazil),
CERN, 
IHEP and LZU (China),
IN2P3 (France), 
KIT and DESY (Germany), 
INFN (Italy), 
SURF (Netherlands),
Polish WLCG (Poland),
IFIN-HH (Romania), 
%RRCKI and Yandex LLC (Russia), 
PIC (Spain), CSCS (Switzerland), 
and GridPP (United Kingdom).
%%%%%%%%%%%%%%%%%%%%%%%%%%%%%%%%%%%%%%%%%%
We are indebted to the communities behind the multiple open-source
software packages on which we depend.
%%%%%%%%%%%%%%%%%%%%%%%%%%%%%%%%%%%%%%%%%%
Individual groups or members have received support from
%ARC and ARDC (Australia); % moved to national 16/01
Key Research Program of Frontier Sciences of CAS, CAS PIFI, CAS CCEPP, 
%Fundamental Research Funds for the Central Universities,  and Sci.\
%\& Tech.\ Program of Guangzhou (China); Removed 24/11/25
Minciencias (Colombia);
EPLANET, Marie Sk\l{}odowska-Curie Actions, ERC and NextGenerationEU (European Union);
A*MIDEX, ANR, IPhU and Labex P2IO, and R\'{e}gion Auvergne-Rh\^{o}ne-Alpes (France);
%RFBR, RSF and Yandex LLC (Russia);
Alexander-von-Humboldt Foundation (Germany);
ICSC (Italy); 
%GVA, XuntaGal, GENCAT, Inditex, InTalent and Prog.~Atracci\'on Talento, CM (Spain);
Severo Ochoa and Mar\'ia de Maeztu Units of Excellence, GVA, XuntaGal, GENCAT, InTalent-Inditex and Prog.~Atracci\'on Talento CM (Spain);
SRC (Sweden);
the Leverhulme Trust, the Royal Society and UKRI (United Kingdom).

\clearpage
% ===============================================================================
% Purpose: appendix to the standard template: standard symbol alises from Ulrik
% Author: Tomasz Skwarnicki
% Created on: 2009-09-24
% ===============================================================================
%{\noindent\normalfont\bfseries\Large Appendices}

\appendix
\section*{End Matter}

\begin{figure}[p]
\centering
\includegraphics[width=0.75\linewidth]{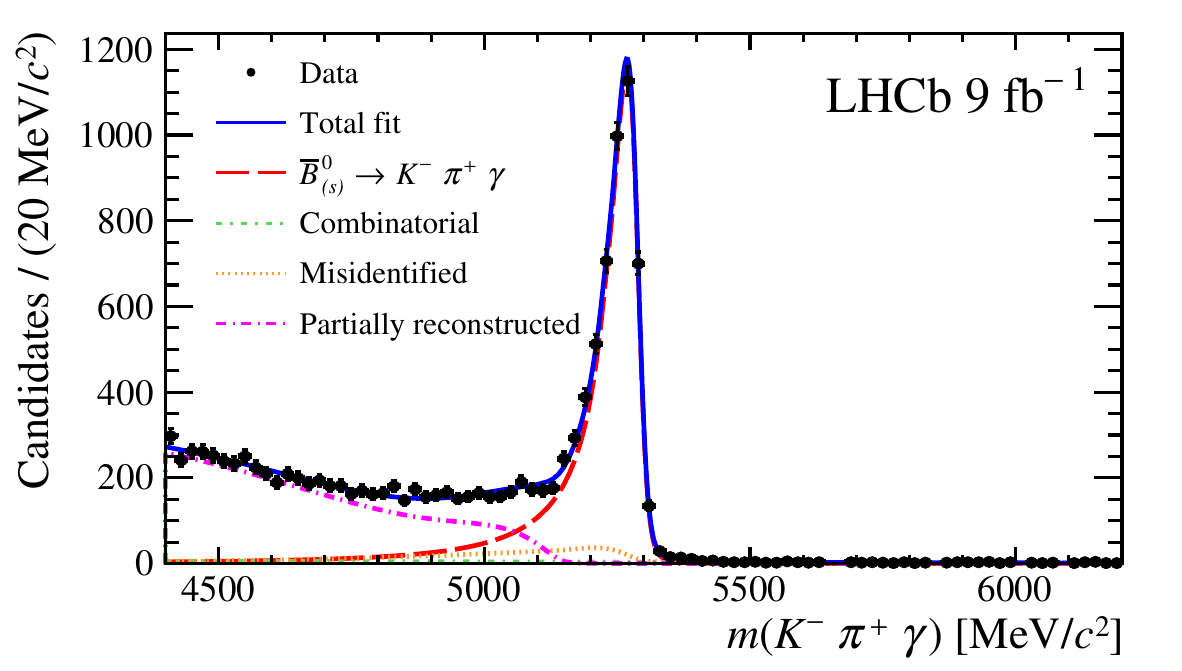}
\includegraphics[width=0.75\linewidth]{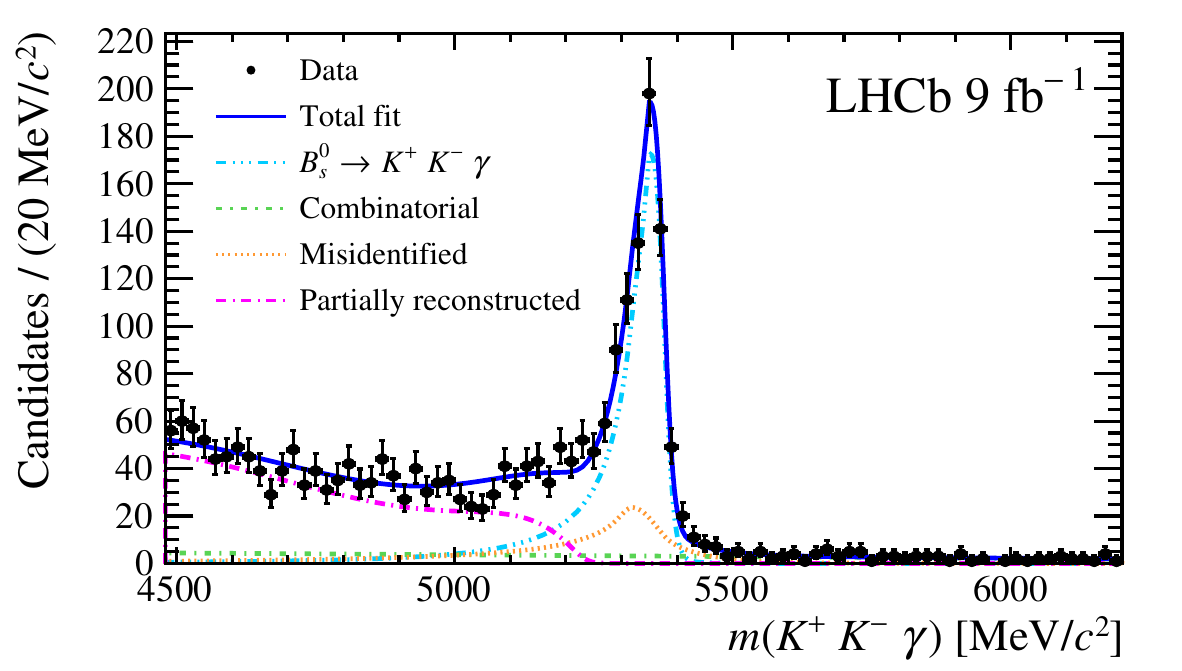}
\includegraphics[width=0.75\linewidth]{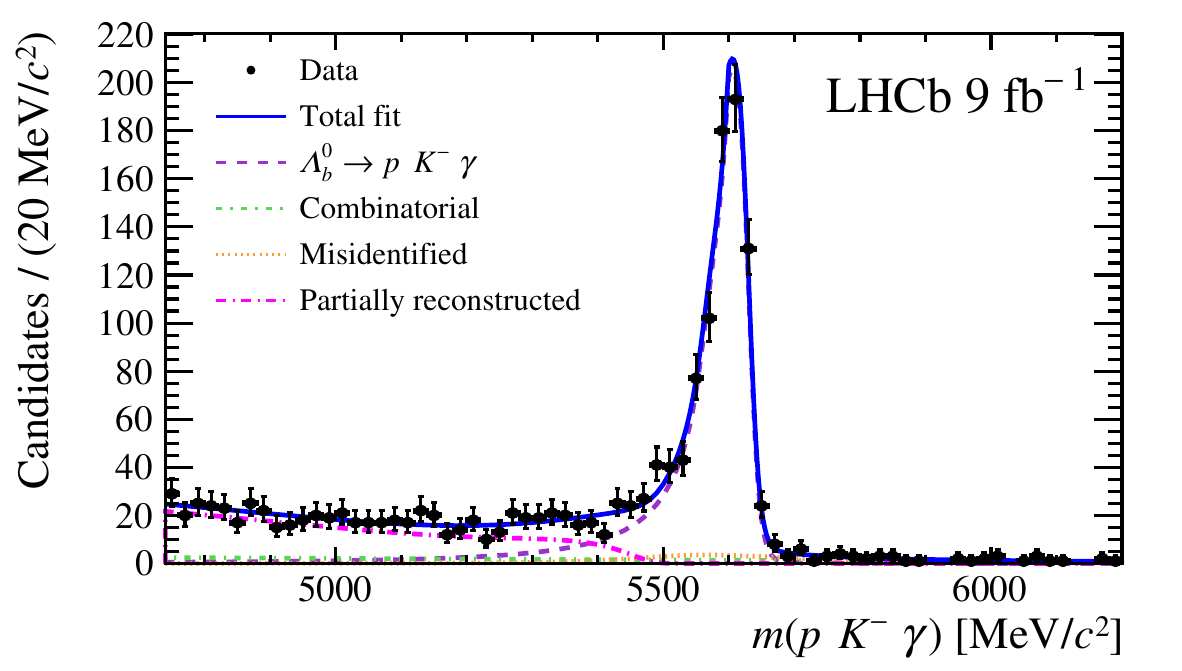}
\caption{Mass distribution of selected (top) \kern-0.4em\shortstack{{\miniscule (\rule[.5ex]{1.25em}{.18mm})}
  \\ [-.7ex] {\kern0.4em $B_{(s)}^0$}}$\rightarrow K^-\pi^+\gamma$, (middle) $B_s^0\rightarrow K^+K^-\gamma$ and (bottom) $\Lambda_b^0 \rightarrow pK^-\gamma$ candidates, obtained by summing the four data samples considered in the analysis.
The solid curve represents the sum of the fit functions from the individual samples, as described in the text.
To improve visibility, the fit functions for sources of misidentified ($\kern 0.18em\overline{\kern -0.18em B}{}^0\rightarrow K^- \pi^+\gamma$, $B_s^0\rightarrow K^+K^-\gamma$, $\Lambda_b^0 \rightarrow pK^-\gamma$, $\B_{(s)}^0 \rightarrow K^-\pi^+\pi^0$, $\B_{(s)}^0 \rightarrow K^-\pi^+\eta$, $\B_{(s)}^0 \rightarrow K^-\pi^+e^+e^-$, $\B_{(s)}^0 \rightarrow K^-\pi^+\pi^-\pi^+$) and partially reconstructed ($\B_{(s)} \rightarrow K^-\pi^+\pi(\pi)\gamma$) backgrounds have been summed.}
\label{fig:datafit_allFS}
\end{figure}

The yields of \bskkg and \lbpkg decays misidentified as \kpig candidates are evaluated using an extended unbinned maximum-likelihood fit performed simultaneously to the \kpig, \kkg and \pkg final states.
To reduce the dependence of the misidentified yield evaluation on the dihadron spectrum of these decays, the requirements on \mkpi are also applied on the invariant mass of the $\Kp\Km$ and $p\Km$ combinations when the pion mass hypothesis is applied to one of the kaons in the \bskkg decay or to the proton.
The yields of the misidentified decays are parametrised as their yield measured in the correct final states multiplied by the selection efficiency ratio.
Selection efficiencies are derived using simulated samples of \mbox{\decay{H_b}{\left[\Km h^+\right]_{X}\gamma}} decays, where $H_b$ is a \bquark-hadron, $h^+$ is a charged pion, kaon or proton and $X$ denotes a resonant state with a large fit fraction.
The efficiencies for the \mbox{\decay{\Bzb}{\left[\Km\pip\right]_X\gamma}} and \mbox{\decay{\Bs}{\left[\Kp\Km\right]_X\gamma}} decays are evaluated as weighted averages of all considered resonant states, where the efficiency for each decay mode is weighted by its fit fraction~\cite{PDG2024,LHCb-PAPER-2024-002}.
The uncertainties on the fit fractions are enlarged by the deviation of the summed fit fractions from unity to account for nonresonant, interference, and neglected contributions.
Since the \mbox{\decay{\Lb}{\left[p\Km\right]_X\gamma}} decay includes a very large number of interfering resonances~\cite{LHCb-PAPER-2023-036}, four resonances of different masses are considered and each assigned equal weight in the average.
When evaluating the systematic uncertainties associated with external inputs, pseudoexperiments are generated using an alternative model that includes only one of the four resonances in turn.
Figure~\ref{fig:datafit_allFS} shows the invariant-mass distribution of candidates selected in each of the three final states summed over the four analysed datasets, as well as the summed fit models.

\newpage
% \section*{PRL Justification}

% \begin{verbatim}
% This Letter presents the first search and evidence for the rare 
% Bs->KpiGamma decay and a measurement of its branching fraction. 
% It adds to the small set of measured b->dGamma radiative penguin 
% decays, but is the first such exclusive decay that will permit a determination 
% of the Cabibbo--Kobayashi--Maskawa matrix element ratio |Vtd/Vts|
% with only small reliance on non-perturbative calculations. A 
% discrepancy with determinations from B-meson oscillations would 
% indicate contributions from physics beyond the Standard Model.
% Therefore this measurement opens a new door for theoretically
% clean tests of the Standard Model.
% \end{verbatim}

% \section*{Length}
% \begin{verbatim}
%     3276 + [4 * (20 + 150/1.77)] + [20 + 150/8.36] + [26 + (15+8)*6.5] + [3*16]
%     = 3956
% \end{verbatim}

\addcontentsline{toc}{section}{References}
%\setboolean{inbibliography}{true}
\bibliographystyle{LHCb/LHCb}
\bibliography{main,LHCb/standard,LHCb/LHCb-PAPER,LHCb/LHCb-CONF,LHCb/LHCb-DP,LHCb/LHCb-TDR}

\ifx\mcitethebibliography\mciteundefinedmacro
\PackageError{LHCb.bst}{mciteplus.sty has not been loaded}
{This bibstyle requires the use of the mciteplus package.}\fi
\providecommand{\href}[2]{#2}
\begin{mcitethebibliography}{10}
\mciteSetBstSublistMode{n}
\mciteSetBstMaxWidthForm{subitem}{\alph{mcitesubitemcount})}
\mciteSetBstSublistLabelBeginEnd{\mcitemaxwidthsubitemform\space}
{\relax}{\relax}

\bibitem{Li:2013vlx}
X.-Q. Li, Y.-D. Yang, and X.-B. Yuan, \ifthenelse{\boolean{articletitles}}{\emph{{Exclusive radiative B-meson decays within minimal flavor-violating two-Higgs-doublet models}}, }{}\href{https://doi.org/10.1103/PhysRevD.89.054024}{Phys.\ Rev.\  \textbf{D89} (2014) 054024}, \href{http://arxiv.org/abs/1311.2786}{{\normalfont\ttfamily arXiv:1311.2786}}\relax
\mciteBstWouldAddEndPuncttrue
\mciteSetBstMidEndSepPunct{\mcitedefaultmidpunct}
{\mcitedefaultendpunct}{\mcitedefaultseppunct}\relax
\EndOfBibitem
\bibitem{Xiao:2003vq}
Z.~Xiao and C.~Zhuang, \ifthenelse{\boolean{articletitles}}{\emph{{Exclusive $B\to (K^{*}, \rho) \gamma$ decays in the general two Higgs doublet models}}, }{}\href{https://doi.org/10.1140/epjc/s2004-01592-0}{Eur.\ Phys.\ J.\  \textbf{C33} (2004) 349}, \href{http://arxiv.org/abs/hep-ph/0310097}{{\normalfont\ttfamily arXiv:hep-ph/0310097}}\relax
\mciteBstWouldAddEndPuncttrue
\mciteSetBstMidEndSepPunct{\mcitedefaultmidpunct}
{\mcitedefaultendpunct}{\mcitedefaultseppunct}\relax
\EndOfBibitem
\bibitem{Ali:2002kw}
A.~Ali and E.~Lunghi, \ifthenelse{\boolean{articletitles}}{\emph{{Implications of $B \to \rho \gamma$ measurements in the standard model and supersymmetric theories}}, }{}\href{https://doi.org/10.1140/epjc/s2002-01053-x}{Eur.\ Phys.\ J.\  \textbf{C26} (2002) 195}, \href{http://arxiv.org/abs/hep-ph/0206242}{{\normalfont\ttfamily arXiv:hep-ph/0206242}}\relax
\mciteBstWouldAddEndPuncttrue
\mciteSetBstMidEndSepPunct{\mcitedefaultmidpunct}
{\mcitedefaultendpunct}{\mcitedefaultseppunct}\relax
\EndOfBibitem
\bibitem{Arhrib:2001gr}
A.~Arhrib, C.-K. Chua, and W.-S. Hou, \ifthenelse{\boolean{articletitles}}{\emph{{Supersymmetric model contributions to $B^0_{d} - \bar{B}^0_d$ mixing and $B \to \pi \pi$, $\rho \gamma$ decays}}, }{}\href{https://doi.org/10.1007/s100520100758}{Eur.\ Phys.\ J.\  \textbf{C21} (2001) 567}, \href{http://arxiv.org/abs/hep-ph/0104122}{{\normalfont\ttfamily arXiv:hep-ph/0104122}}\relax
\mciteBstWouldAddEndPuncttrue
\mciteSetBstMidEndSepPunct{\mcitedefaultmidpunct}
{\mcitedefaultendpunct}{\mcitedefaultseppunct}\relax
\EndOfBibitem
\bibitem{LHCb-PAPER-2019-015}
LHCb collaboration, R.~Aaij {\em et~al.}, \ifthenelse{\boolean{articletitles}}{\emph{{Measurement of \CP-violating and mixing-induced observables in \mbox{\decay{\Bs}{\phi \gamma}} decays}}, }{}\href{https://doi.org/10.1103/PhysRevLett.123.081802}{Phys.\ Rev.\ Lett.\  \textbf{123} (2019) 081802}, \href{http://arxiv.org/abs/1905.06284}{{\normalfont\ttfamily arXiv:1905.06284}}\relax
\mciteBstWouldAddEndPuncttrue
\mciteSetBstMidEndSepPunct{\mcitedefaultmidpunct}
{\mcitedefaultendpunct}{\mcitedefaultseppunct}\relax
\EndOfBibitem
\bibitem{LHCb-PAPER-2021-030}
LHCb collaboration, R.~Aaij {\em et~al.}, \ifthenelse{\boolean{articletitles}}{\emph{{Measurement of the photon polarization in $\Lb \to \Lz \gamma$ decays}}, }{}\href{https://doi.org/10.1103/PhysRevD.105.L051104}{Phys.\ Rev.\  \textbf{D105} (2022) 051104}, \href{http://arxiv.org/abs/2111.10194}{{\normalfont\ttfamily arXiv:2111.10194}}\relax
\mciteBstWouldAddEndPuncttrue
\mciteSetBstMidEndSepPunct{\mcitedefaultmidpunct}
{\mcitedefaultendpunct}{\mcitedefaultseppunct}\relax
\EndOfBibitem
\bibitem{LHCb-PAPER-2020-020}
LHCb collaboration, R.~Aaij {\em et~al.}, \ifthenelse{\boolean{articletitles}}{\emph{{Strong constraints on the $b \to s\gamma$ photon polarisation from $B^0 \to K^{\ast 0} e^+ e^-$ decays}}, }{}\href{https://doi.org/10.1007/JHEP12(2020)081}{JHEP \textbf{12} (2020) 081}, \href{http://arxiv.org/abs/2010.06011}{{\normalfont\ttfamily arXiv:2010.06011}}\relax
\mciteBstWouldAddEndPuncttrue
\mciteSetBstMidEndSepPunct{\mcitedefaultmidpunct}
{\mcitedefaultendpunct}{\mcitedefaultseppunct}\relax
\EndOfBibitem
\bibitem{LHCb-PAPER-2024-030}
LHCb collaboration, R.~Aaij {\em et~al.}, \ifthenelse{\boolean{articletitles}}{\emph{{Constraints on the photon polarisation in $b \to s \gamma$ transitions using $B_s^0\to \phi e^+ e^-$ decays}}, }{}\href{https://doi.org/10.1007/JHEP03(2025)047}{{JHEP} \textbf{03} (2025) 047}, \href{http://arxiv.org/abs/2411.10219}{{\normalfont\ttfamily arXiv:2411.10219}}\relax
\mciteBstWouldAddEndPuncttrue
\mciteSetBstMidEndSepPunct{\mcitedefaultmidpunct}
{\mcitedefaultendpunct}{\mcitedefaultseppunct}\relax
\EndOfBibitem
\bibitem{LHCb-PAPER-2012-019}
LHCb collaboration, R.~Aaij {\em et~al.}, \ifthenelse{\boolean{articletitles}}{\emph{{Measurement of the ratio of branching fractions \mbox{$\BF(\decay{\Bz}{\Kstarz\g}) / \BF(\decay{\Bs}{\phiz\g})$} and the direct \CP asymmetry in \mbox{\decay{\Bz}{\Kstarz\g}}}}, }{}\href{https://doi.org/10.1016/j.nuclphysb.2012.09.013}{Nucl.\ Phys.\  \textbf{B867} (2013) 1}, \href{http://arxiv.org/abs/1209.0313}{{\normalfont\ttfamily arXiv:1209.0313}}\relax
\mciteBstWouldAddEndPuncttrue
\mciteSetBstMidEndSepPunct{\mcitedefaultmidpunct}
{\mcitedefaultendpunct}{\mcitedefaultseppunct}\relax
\EndOfBibitem
\bibitem{Belle-II:2024kbc}
Belle II collaboration, I.~Adachi {\em et~al.}, \ifthenelse{\boolean{articletitles}}{\emph{{Measurement of $B\to K^{*}(892)\gamma$ decays at Belle II}}, }{}\href{https://doi.org/10.1007/JHEP09(2025)024}{JHEP \textbf{09} (2025) 024}, \href{http://arxiv.org/abs/2411.10127}{{\normalfont\ttfamily arXiv:2411.10127}}\relax
\mciteBstWouldAddEndPuncttrue
\mciteSetBstMidEndSepPunct{\mcitedefaultmidpunct}
{\mcitedefaultendpunct}{\mcitedefaultseppunct}\relax
\EndOfBibitem
\bibitem{BaBar:2015chw}
BaBar collaboration, P.~del Amo~Sanchez {\em et~al.}, \ifthenelse{\boolean{articletitles}}{\emph{{Time-dependent analysis of $B^0 \to {{K^0_{S}}} \pi^- \pi^+ \gamma$ decays and studies of the $K^+\pi^-\pi^+$ system in $B^+ \to K^+ \pi^- \pi^+ \gamma$ decays}}, }{}\href{https://doi.org/10.1103/PhysRevD.93.052013}{Phys.\ Rev.\  \textbf{D93} (2016) 052013}, \href{http://arxiv.org/abs/1512.03579}{{\normalfont\ttfamily arXiv:1512.03579}}\relax
\mciteBstWouldAddEndPuncttrue
\mciteSetBstMidEndSepPunct{\mcitedefaultmidpunct}
{\mcitedefaultendpunct}{\mcitedefaultseppunct}\relax
\EndOfBibitem
\bibitem{Paul:2016urs}
A.~Paul and D.~M. Straub, \ifthenelse{\boolean{articletitles}}{\emph{{Constraints on new physics from radiative $B$ decays}}, }{}\href{https://doi.org/10.1007/JHEP04(2017)027}{JHEP \textbf{04} (2017) 027}, \href{http://arxiv.org/abs/1608.02556}{{\normalfont\ttfamily arXiv:1608.02556}}\relax
\mciteBstWouldAddEndPuncttrue
\mciteSetBstMidEndSepPunct{\mcitedefaultmidpunct}
{\mcitedefaultendpunct}{\mcitedefaultseppunct}\relax
\EndOfBibitem
\bibitem{LHCb-PAPER-2025-017}
LHCb collaboration, R.~Aaij {\em et~al.}, \ifthenelse{\boolean{articletitles}}{\emph{{Measurement of the $B^0 \to \rho^0 \gamma$ branching fraction}}, }{}\href{http://arxiv.org/abs/2507.14401}{{\normalfont\ttfamily arXiv:2507.14401}}, {Submitted to JHEP}\relax
\mciteBstWouldAddEndPuncttrue
\mciteSetBstMidEndSepPunct{\mcitedefaultmidpunct}
{\mcitedefaultendpunct}{\mcitedefaultseppunct}\relax
\EndOfBibitem
\bibitem{Belle:2024qhe}
{Belle and Belle II collaborations}, I.~Adachi {\em et~al.}, \ifthenelse{\boolean{articletitles}}{\emph{{Measurement of branching fractions, \CP asymmetry, and isospin asymmetry for $B \to \rho\gamma$ decays using Belle and Belle II data}}, }{}\href{https://doi.org/10.1103/PhysRevD.111.L071103}{Phys.\ Rev.\  \textbf{D111} (2025) L071103}, \href{http://arxiv.org/abs/2407.08984}{{\normalfont\ttfamily arXiv:2407.08984}}\relax
\mciteBstWouldAddEndPuncttrue
\mciteSetBstMidEndSepPunct{\mcitedefaultmidpunct}
{\mcitedefaultendpunct}{\mcitedefaultseppunct}\relax
\EndOfBibitem
\bibitem{Belle:2008imr}
Belle collaboration, N.~Taniguchi {\em et~al.}, \ifthenelse{\boolean{articletitles}}{\emph{{Measurement of branching fractions, isospin and \CP-violating asymmetries for exclusive $b \to d \gamma$ modes}}, }{}\href{https://doi.org/10.1103/PhysRevLett.101.111801}{Phys.\ Rev.\ Lett.\  \textbf{101} (2008) 111801}, Erratum \href{https://doi.org/10.1103/PhysRevLett.101.129904}{ibid.\   \textbf{101} (2008) 129904}, \href{http://arxiv.org/abs/0804.4770}{{\normalfont\ttfamily arXiv:0804.4770}}\relax
\mciteBstWouldAddEndPuncttrue
\mciteSetBstMidEndSepPunct{\mcitedefaultmidpunct}
{\mcitedefaultendpunct}{\mcitedefaultseppunct}\relax
\EndOfBibitem
\bibitem{BaBar:2008txc}
BaBar collaboration, B.~Aubert {\em et~al.}, \ifthenelse{\boolean{articletitles}}{\emph{{Measurements of branching fractions for $B^{+} \to \rho^{+} \gamma$, $B^0 \to \rho^0 \gamma$, and $B^0 \to \omega \gamma$}}, }{}\href{https://doi.org/10.1103/PhysRevD.78.112001}{Phys.\ Rev.\  \textbf{D78} (2008) 112001}, \href{http://arxiv.org/abs/0808.1379}{{\normalfont\ttfamily arXiv:0808.1379}}\relax
\mciteBstWouldAddEndPuncttrue
\mciteSetBstMidEndSepPunct{\mcitedefaultmidpunct}
{\mcitedefaultendpunct}{\mcitedefaultseppunct}\relax
\EndOfBibitem
\bibitem{BaBar:2010vgu}
BaBar collaboration, P.~del Amo~Sanchez {\em et~al.}, \ifthenelse{\boolean{articletitles}}{\emph{{Study of $B \to X\gamma$ decays and determination of $|V_{td}/V_{ts}|$}}, }{}\href{https://doi.org/10.1103/PhysRevD.82.051101}{Phys.\ Rev.\  \textbf{D82} (2010) 051101}, \href{http://arxiv.org/abs/1005.4087}{{\normalfont\ttfamily arXiv:1005.4087}}\relax
\mciteBstWouldAddEndPuncttrue
\mciteSetBstMidEndSepPunct{\mcitedefaultmidpunct}
{\mcitedefaultendpunct}{\mcitedefaultseppunct}\relax
\EndOfBibitem
\bibitem{Ball:2006nr}
P.~Ball and R.~Zwicky, \ifthenelse{\boolean{articletitles}}{\emph{{$|V_{td} / V_{ts}|$ from $B \to V \gamma$}}, }{}\href{https://doi.org/10.1088/1126-6708/2006/04/046}{JHEP \textbf{04} (2006) 046}, \href{http://arxiv.org/abs/hep-ph/0603232}{{\normalfont\ttfamily arXiv:hep-ph/0603232}}\relax
\mciteBstWouldAddEndPuncttrue
\mciteSetBstMidEndSepPunct{\mcitedefaultmidpunct}
{\mcitedefaultendpunct}{\mcitedefaultseppunct}\relax
\EndOfBibitem
\bibitem{Ball:2006eu}
P.~Ball, G.~W. Jones, and R.~Zwicky, \ifthenelse{\boolean{articletitles}}{\emph{{$B \to V \gamma$ beyond QCD factorisation}}, }{}\href{https://doi.org/10.1103/PhysRevD.75.054004}{Phys.\ Rev.\  \textbf{D75} (2007) 054004}, \href{http://arxiv.org/abs/hep-ph/0612081}{{\normalfont\ttfamily arXiv:hep-ph/0612081}}\relax
\mciteBstWouldAddEndPuncttrue
\mciteSetBstMidEndSepPunct{\mcitedefaultmidpunct}
{\mcitedefaultendpunct}{\mcitedefaultseppunct}\relax
\EndOfBibitem
\bibitem{LHCb-PAPER-2021-005}
LHCb collaboration, R.~Aaij {\em et~al.}, \ifthenelse{\boolean{articletitles}}{\emph{{Precise determination of the $\Bs$-$\Bsb$ oscillation frequency}}, }{}\href{https://doi.org/10.1038/s41567-021-01394-x}{Nature Physics \textbf{18} (2022) 1}, \href{http://arxiv.org/abs/2104.04421}{{\normalfont\ttfamily arXiv:2104.04421}}\relax
\mciteBstWouldAddEndPuncttrue
\mciteSetBstMidEndSepPunct{\mcitedefaultmidpunct}
{\mcitedefaultendpunct}{\mcitedefaultseppunct}\relax
\EndOfBibitem
\bibitem{HFLAV23}
Heavy Flavor Averaging Group, S.~Banerjee {\em et~al.}, \ifthenelse{\boolean{articletitles}}{\emph{{Averages of $b$-hadron, $c$-hadron, and $\tau$-lepton properties as of 2023}}, }{}\href{http://arxiv.org/abs/2411.18639}{{\normalfont\ttfamily arXiv:2411.18639}}, {updated results and plots available at \href{https://hflav.web.cern.ch}{{\texttt{https://hflav.web.cern.ch}}}}\relax
\mciteBstWouldAddEndPuncttrue
\mciteSetBstMidEndSepPunct{\mcitedefaultmidpunct}
{\mcitedefaultendpunct}{\mcitedefaultseppunct}\relax
\EndOfBibitem
\bibitem{CKMfitter2005}
CKMfitter group, J.~Charles {\em et~al.}, \ifthenelse{\boolean{articletitles}}{\emph{{\CP violation and the CKM matrix: Assessing the impact of the asymmetric $B$ factories}}, }{}\href{https://doi.org/10.1140/epjc/s2005-02169-1}{Eur.\ Phys.\ J.\  \textbf{C41} (2005) 1}, \href{http://arxiv.org/abs/hep-ph/0406184}{{\normalfont\ttfamily arXiv:hep-ph/0406184}}, {updated results and plots available at \href{http://ckmfitter.in2p3.fr/}{{\texttt{http://ckmfitter.in2p3.fr/}}}}\relax
\mciteBstWouldAddEndPuncttrue
\mciteSetBstMidEndSepPunct{\mcitedefaultmidpunct}
{\mcitedefaultendpunct}{\mcitedefaultseppunct}\relax
\EndOfBibitem
\bibitem{CKMfitter2015}
CKMfitter group, J.~Charles {\em et~al.}, \ifthenelse{\boolean{articletitles}}{\emph{{Current status of the standard model CKM fit and constraints on \hbox{$\Delta F=2$} new physics}}, }{}\href{https://doi.org/10.1103/PhysRevD.91.073007}{Phys.\ Rev.\  \textbf{D91} (2015) 073007}, \href{http://arxiv.org/abs/1501.05013}{{\normalfont\ttfamily arXiv:1501.05013}}, {updated results and plots available at \href{http://ckmfitter.in2p3.fr/}{{\texttt{http://ckmfitter.in2p3.fr/}}}}\relax
\mciteBstWouldAddEndPuncttrue
\mciteSetBstMidEndSepPunct{\mcitedefaultmidpunct}
{\mcitedefaultendpunct}{\mcitedefaultseppunct}\relax
\EndOfBibitem
\bibitem{PDG2024}
Particle Data Group, S.~Navas {\em et~al.}, \ifthenelse{\boolean{articletitles}}{\emph{{\href{http://pdg.lbl.gov/}{Review of particle physics}}}, }{}\href{https://doi.org/10.1103/PhysRevD.110.030001}{Phys.\ Rev.\  \textbf{D110} (2024) 030001}\relax
\mciteBstWouldAddEndPuncttrue
\mciteSetBstMidEndSepPunct{\mcitedefaultmidpunct}
{\mcitedefaultendpunct}{\mcitedefaultseppunct}\relax
\EndOfBibitem
\bibitem{LHCb-DP-2008-001}
LHCb collaboration, A.~A. Alves~Jr.\ {\em et~al.}, \ifthenelse{\boolean{articletitles}}{\emph{{The \lhcb detector at the LHC}}, }{}\href{https://doi.org/10.1088/1748-0221/3/08/S08005}{JINST \textbf{3} (2008) S08005}\relax
\mciteBstWouldAddEndPuncttrue
\mciteSetBstMidEndSepPunct{\mcitedefaultmidpunct}
{\mcitedefaultendpunct}{\mcitedefaultseppunct}\relax
\EndOfBibitem
\bibitem{LHCb-DP-2014-002}
LHCb collaboration, R.~Aaij {\em et~al.}, \ifthenelse{\boolean{articletitles}}{\emph{{LHCb detector performance}}, }{}\href{https://doi.org/10.1142/S0217751X15300227}{Int.\ J.\ Mod.\ Phys.\  \textbf{A30} (2015) 1530022}, \href{http://arxiv.org/abs/1412.6352}{{\normalfont\ttfamily arXiv:1412.6352}}\relax
\mciteBstWouldAddEndPuncttrue
\mciteSetBstMidEndSepPunct{\mcitedefaultmidpunct}
{\mcitedefaultendpunct}{\mcitedefaultseppunct}\relax
\EndOfBibitem
\bibitem{LHCb-DP-2014-001}
R.~Aaij {\em et~al.}, \ifthenelse{\boolean{articletitles}}{\emph{{Performance of the LHCb Vertex Locator}}, }{}\href{https://doi.org/10.1088/1748-0221/9/09/P09007}{JINST \textbf{9} (2014) P09007}, \href{http://arxiv.org/abs/1405.7808}{{\normalfont\ttfamily arXiv:1405.7808}}\relax
\mciteBstWouldAddEndPuncttrue
\mciteSetBstMidEndSepPunct{\mcitedefaultmidpunct}
{\mcitedefaultendpunct}{\mcitedefaultseppunct}\relax
\EndOfBibitem
\bibitem{LHCb-DP-2013-003}
R.~Arink {\em et~al.}, \ifthenelse{\boolean{articletitles}}{\emph{{Performance of the LHCb Outer Tracker}}, }{}\href{https://doi.org/10.1088/1748-0221/9/01/P01002}{JINST \textbf{9} (2014) P01002}, \href{http://arxiv.org/abs/1311.3893}{{\normalfont\ttfamily arXiv:1311.3893}}\relax
\mciteBstWouldAddEndPuncttrue
\mciteSetBstMidEndSepPunct{\mcitedefaultmidpunct}
{\mcitedefaultendpunct}{\mcitedefaultseppunct}\relax
\EndOfBibitem
\bibitem{LHCb-DP-2017-001}
P.~d'Argent {\em et~al.}, \ifthenelse{\boolean{articletitles}}{\emph{{Improved performance of the LHCb Outer Tracker in LHC Run 2}}, }{}\href{https://doi.org/10.1088/1748-0221/12/11/P11016}{JINST \textbf{12} (2017) P11016}, \href{http://arxiv.org/abs/1708.00819}{{\normalfont\ttfamily arXiv:1708.00819}}\relax
\mciteBstWouldAddEndPuncttrue
\mciteSetBstMidEndSepPunct{\mcitedefaultmidpunct}
{\mcitedefaultendpunct}{\mcitedefaultseppunct}\relax
\EndOfBibitem
\bibitem{LHCb-DP-2012-003}
M.~Adinolfi {\em et~al.}, \ifthenelse{\boolean{articletitles}}{\emph{{Performance of the \lhcb RICH detector at the LHC}}, }{}\href{https://doi.org/10.1140/epjc/s10052-013-2431-9}{Eur.\ Phys.\ J.\  \textbf{C73} (2013) 2431}, \href{http://arxiv.org/abs/1211.6759}{{\normalfont\ttfamily arXiv:1211.6759}}\relax
\mciteBstWouldAddEndPuncttrue
\mciteSetBstMidEndSepPunct{\mcitedefaultmidpunct}
{\mcitedefaultendpunct}{\mcitedefaultseppunct}\relax
\EndOfBibitem
\bibitem{LHCb-DP-2020-001}
C.~Abell\'an~Beteta {\em et~al.}, \ifthenelse{\boolean{articletitles}}{\emph{{Calibration and performance of the LHCb calorimeters in Run 1 and 2 at the LHC}}, }{}\href{http://arxiv.org/abs/2008.11556}{{\normalfont\ttfamily arXiv:2008.11556}}\relax
\mciteBstWouldAddEndPuncttrue
\mciteSetBstMidEndSepPunct{\mcitedefaultmidpunct}
{\mcitedefaultendpunct}{\mcitedefaultseppunct}\relax
\EndOfBibitem
\bibitem{LHCb-DP-2012-004}
R.~Aaij {\em et~al.}, \ifthenelse{\boolean{articletitles}}{\emph{{The \lhcb trigger and its performance in 2011}}, }{}\href{https://doi.org/10.1088/1748-0221/8/04/P04022}{JINST \textbf{8} (2013) P04022}, \href{http://arxiv.org/abs/1211.3055}{{\normalfont\ttfamily arXiv:1211.3055}}\relax
\mciteBstWouldAddEndPuncttrue
\mciteSetBstMidEndSepPunct{\mcitedefaultmidpunct}
{\mcitedefaultendpunct}{\mcitedefaultseppunct}\relax
\EndOfBibitem
\bibitem{LHCb-DP-2019-001}
R.~Aaij {\em et~al.}, \ifthenelse{\boolean{articletitles}}{\emph{{Design and performance of the LHCb trigger and full real-time reconstruction in Run 2 of the LHC}}, }{}\href{https://doi.org/10.1088/1748-0221/14/04/P04013}{JINST \textbf{14} (2019) P04013}, \href{http://arxiv.org/abs/1812.10790}{{\normalfont\ttfamily arXiv:1812.10790}}\relax
\mciteBstWouldAddEndPuncttrue
\mciteSetBstMidEndSepPunct{\mcitedefaultmidpunct}
{\mcitedefaultendpunct}{\mcitedefaultseppunct}\relax
\EndOfBibitem
\bibitem{BBDT}
V.~V. Gligorov and M.~Williams, \ifthenelse{\boolean{articletitles}}{\emph{{Efficient, reliable and fast high-level triggering using a bonsai boosted decision tree}}, }{}\href{https://doi.org/10.1088/1748-0221/8/02/P02013}{JINST \textbf{8} (2013) P02013}, \href{http://arxiv.org/abs/1210.6861}{{\normalfont\ttfamily arXiv:1210.6861}}\relax
\mciteBstWouldAddEndPuncttrue
\mciteSetBstMidEndSepPunct{\mcitedefaultmidpunct}
{\mcitedefaultendpunct}{\mcitedefaultseppunct}\relax
\EndOfBibitem
\bibitem{LHCb-PROC-2015-018}
T.~Likhomanenko {\em et~al.}, \ifthenelse{\boolean{articletitles}}{\emph{{LHCb topological trigger reoptimization}}, }{}\href{https://doi.org/10.1088/1742-6596/664/8/082025}{J.\ Phys.\ Conf.\ Ser.\  \textbf{664} (2015) 082025}, \href{http://arxiv.org/abs/1510.00572}{{\normalfont\ttfamily arXiv:1510.00572}}\relax
\mciteBstWouldAddEndPuncttrue
\mciteSetBstMidEndSepPunct{\mcitedefaultmidpunct}
{\mcitedefaultendpunct}{\mcitedefaultseppunct}\relax
\EndOfBibitem
\bibitem{LHCb-PUB-2012-002}
A.~Puig, \ifthenelse{\boolean{articletitles}}{\emph{{The HLT2 radiative topological lines}}, }{} \href{http://cdsweb.cern.ch/search?p=LHCb-PUB-2012-002, CERN-LHCb-PUB-2012-002&f=reportnumber&action_search=Search&c=LHCb+Notes} {LHCb-PUB-2012-002, CERN-LHCb-PUB-2012-002}, 2012\relax
\mciteBstWouldAddEndPuncttrue
\mciteSetBstMidEndSepPunct{\mcitedefaultmidpunct}
{\mcitedefaultendpunct}{\mcitedefaultseppunct}\relax
\EndOfBibitem
\bibitem{Sjostrand:2007gs}
T.~Sj\"{o}strand, S.~Mrenna, and P.~Skands, \ifthenelse{\boolean{articletitles}}{\emph{{A brief introduction to PYTHIA 8.1}}, }{}\href{https://doi.org/10.1016/j.cpc.2008.01.036}{Comput.\ Phys.\ Commun.\  \textbf{178} (2008) 852}, \href{http://arxiv.org/abs/0710.3820}{{\normalfont\ttfamily arXiv:0710.3820}}\relax
\mciteBstWouldAddEndPuncttrue
\mciteSetBstMidEndSepPunct{\mcitedefaultmidpunct}
{\mcitedefaultendpunct}{\mcitedefaultseppunct}\relax
\EndOfBibitem
\bibitem{Sjostrand:2006za}
T.~Sj\"{o}strand, S.~Mrenna, and P.~Skands, \ifthenelse{\boolean{articletitles}}{\emph{{PYTHIA 6.4 physics and manual}}, }{}\href{https://doi.org/10.1088/1126-6708/2006/05/026}{JHEP \textbf{05} (2006) 026}, \href{http://arxiv.org/abs/hep-ph/0603175}{{\normalfont\ttfamily arXiv:hep-ph/0603175}}\relax
\mciteBstWouldAddEndPuncttrue
\mciteSetBstMidEndSepPunct{\mcitedefaultmidpunct}
{\mcitedefaultendpunct}{\mcitedefaultseppunct}\relax
\EndOfBibitem
\bibitem{LHCb-PROC-2010-056}
I.~Belyaev {\em et~al.}, \ifthenelse{\boolean{articletitles}}{\emph{{Handling of the generation of primary events in Gauss, the LHCb simulation framework}}, }{}\href{https://doi.org/10.1088/1742-6596/331/3/032047}{J.\ Phys.\ Conf.\ Ser.\  \textbf{331} (2011) 032047}\relax
\mciteBstWouldAddEndPuncttrue
\mciteSetBstMidEndSepPunct{\mcitedefaultmidpunct}
{\mcitedefaultendpunct}{\mcitedefaultseppunct}\relax
\EndOfBibitem
\bibitem{Lange:2001uf}
D.~J. Lange, \ifthenelse{\boolean{articletitles}}{\emph{{The EvtGen particle decay simulation package}}, }{}\href{https://doi.org/10.1016/S0168-9002(01)00089-4}{Nucl.\ Instrum.\ Meth.\  \textbf{A462} (2001) 152}\relax
\mciteBstWouldAddEndPuncttrue
\mciteSetBstMidEndSepPunct{\mcitedefaultmidpunct}
{\mcitedefaultendpunct}{\mcitedefaultseppunct}\relax
\EndOfBibitem
\bibitem{davidson2015photos}
N.~Davidson, T.~Przedzinski, and Z.~Was, \ifthenelse{\boolean{articletitles}}{\emph{{PHOTOS interface in C++: Technical and physics documentation}}, }{}\href{https://doi.org/https://doi.org/10.1016/j.cpc.2015.09.013}{Comput.\ Phys.\ Commun.\  \textbf{199} (2016) 86}, \href{http://arxiv.org/abs/1011.0937}{{\normalfont\ttfamily arXiv:1011.0937}}\relax
\mciteBstWouldAddEndPuncttrue
\mciteSetBstMidEndSepPunct{\mcitedefaultmidpunct}
{\mcitedefaultendpunct}{\mcitedefaultseppunct}\relax
\EndOfBibitem
\bibitem{Allison:2006ve}
Geant4 collaboration, J.~Allison {\em et~al.}, \ifthenelse{\boolean{articletitles}}{\emph{{Geant4 developments and applications}}, }{}\href{https://doi.org/10.1109/TNS.2006.869826}{IEEE Trans.\ Nucl.\ Sci.\  \textbf{53} (2006) 270}\relax
\mciteBstWouldAddEndPuncttrue
\mciteSetBstMidEndSepPunct{\mcitedefaultmidpunct}
{\mcitedefaultendpunct}{\mcitedefaultseppunct}\relax
\EndOfBibitem
\bibitem{Agostinelli:2002hh}
Geant4 collaboration, S.~Agostinelli {\em et~al.}, \ifthenelse{\boolean{articletitles}}{\emph{{Geant4: A simulation toolkit}}, }{}\href{https://doi.org/10.1016/S0168-9002(03)01368-8}{Nucl.\ Instrum.\ Meth.\  \textbf{A506} (2003) 250}\relax
\mciteBstWouldAddEndPuncttrue
\mciteSetBstMidEndSepPunct{\mcitedefaultmidpunct}
{\mcitedefaultendpunct}{\mcitedefaultseppunct}\relax
\EndOfBibitem
\bibitem{LHCb-PROC-2011-006}
M.~Clemencic {\em et~al.}, \ifthenelse{\boolean{articletitles}}{\emph{{The \lhcb simulation application, Gauss: Design, evolution and experience}}, }{}\href{https://doi.org/10.1088/1742-6596/331/3/032023}{J.\ Phys.\ Conf.\ Ser.\  \textbf{331} (2011) 032023}\relax
\mciteBstWouldAddEndPuncttrue
\mciteSetBstMidEndSepPunct{\mcitedefaultmidpunct}
{\mcitedefaultendpunct}{\mcitedefaultseppunct}\relax
\EndOfBibitem
\bibitem{LHCb-DP-2018-001}
R.~Aaij {\em et~al.}, \ifthenelse{\boolean{articletitles}}{\emph{{Selection and processing of calibration samples to measure the particle identification performance of the LHCb experiment in Run 2}}, }{}\href{https://doi.org/10.1140/epjti/s40485-019-0050-z}{Eur.\ Phys.\ J.\ Tech.\ Instr.\  \textbf{6} (2019) 1}, \href{http://arxiv.org/abs/1803.00824}{{\normalfont\ttfamily arXiv:1803.00824}}\relax
\mciteBstWouldAddEndPuncttrue
\mciteSetBstMidEndSepPunct{\mcitedefaultmidpunct}
{\mcitedefaultendpunct}{\mcitedefaultseppunct}\relax
\EndOfBibitem
\bibitem{LHCb-PAPER-2013-028}
LHCb collaboration, R.~Aaij {\em et~al.}, \ifthenelse{\boolean{articletitles}}{\emph{{Measurement of the relative rate of prompt $\chiczero$, $\chicone$ and $\chictwo$ production at \mbox{$\sqs=$7~\tev}}}, }{}\href{https://doi.org/10.1007/JHEP10(2013)115}{JHEP \textbf{10} (2013) 115}, \href{http://arxiv.org/abs/1307.4285}{{\normalfont\ttfamily arXiv:1307.4285}}\relax
\mciteBstWouldAddEndPuncttrue
\mciteSetBstMidEndSepPunct{\mcitedefaultmidpunct}
{\mcitedefaultendpunct}{\mcitedefaultseppunct}\relax
\EndOfBibitem
\bibitem{LHCb-PAPER-2014-040}
LHCb collaboration, R.~Aaij {\em et~al.}, \ifthenelse{\boolean{articletitles}}{\emph{{Measurement of the $\chi_b(3P)$ mass and of the relative rate of $\chi_{b1}(1P)$ and $\chi_{b2}(1P)$ production}}, }{}\href{https://doi.org/10.1007/JHEP10(2014)088}{JHEP \textbf{10} (2014) 088}, \href{http://arxiv.org/abs/1409.1408}{{\normalfont\ttfamily arXiv:1409.1408}}\relax
\mciteBstWouldAddEndPuncttrue
\mciteSetBstMidEndSepPunct{\mcitedefaultmidpunct}
{\mcitedefaultendpunct}{\mcitedefaultseppunct}\relax
\EndOfBibitem
\bibitem{LHCb-PAPER-2015-044}
LHCb collaboration, R.~Aaij {\em et~al.}, \ifthenelse{\boolean{articletitles}}{\emph{{Search for the decays \mbox{\decay{\Bz}{\jpsi\gamma}} and \mbox{\decay{\Bs}{\jpsi\gamma}}}}, }{}\href{https://doi.org/10.1103/PhysRevD.92.112002}{Phys.\ Rev.\  \textbf{D92} (2015) 112002}, \href{http://arxiv.org/abs/1510.04866}{{\normalfont\ttfamily arXiv:1510.04866}}\relax
\mciteBstWouldAddEndPuncttrue
\mciteSetBstMidEndSepPunct{\mcitedefaultmidpunct}
{\mcitedefaultendpunct}{\mcitedefaultseppunct}\relax
\EndOfBibitem
\bibitem{Hulsbergen:2005pu}
W.~D. Hulsbergen, \ifthenelse{\boolean{articletitles}}{\emph{{Decay chain fitting with a Kalman filter}}, }{}\href{https://doi.org/10.1016/j.nima.2005.06.078}{Nucl.\ Instrum.\ Meth.\  \textbf{A552} (2005) 566}, \href{http://arxiv.org/abs/physics/0503191}{{\normalfont\ttfamily arXiv:physics/0503191}}\relax
\mciteBstWouldAddEndPuncttrue
\mciteSetBstMidEndSepPunct{\mcitedefaultmidpunct}
{\mcitedefaultendpunct}{\mcitedefaultseppunct}\relax
\EndOfBibitem
\bibitem{Breiman}
L.~Breiman, J.~H. Friedman, R.~A. Olshen, and C.~J. Stone, {\em Classification and regression trees}, Wadsworth international group, Belmont, California, USA, 1984\relax
\mciteBstWouldAddEndPuncttrue
\mciteSetBstMidEndSepPunct{\mcitedefaultmidpunct}
{\mcitedefaultendpunct}{\mcitedefaultseppunct}\relax
\EndOfBibitem
\bibitem{AdaBoost}
Y.~Freund and R.~E. Schapire, \ifthenelse{\boolean{articletitles}}{\emph{A decision-theoretic generalization of on-line learning and an application to boosting}, }{}\href{https://doi.org/10.1006/jcss.1997.1504}{J.\ Comput.\ Syst.\ Sci.\  \textbf{55} (1997) 119}\relax
\mciteBstWouldAddEndPuncttrue
\mciteSetBstMidEndSepPunct{\mcitedefaultmidpunct}
{\mcitedefaultendpunct}{\mcitedefaultseppunct}\relax
\EndOfBibitem
\bibitem{Hocker:2007ht}
H.~Voss, A.~Hoecker, J.~Stelzer, and F.~Tegenfeldt, \ifthenelse{\boolean{articletitles}}{\emph{{TMVA - Toolkit for Multivariate Data Analysis with ROOT}}, }{}\href{https://doi.org/10.22323/1.050.0040}{PoS \textbf{ACAT} (2007) 040}\relax
\mciteBstWouldAddEndPuncttrue
\mciteSetBstMidEndSepPunct{\mcitedefaultmidpunct}
{\mcitedefaultendpunct}{\mcitedefaultseppunct}\relax
\EndOfBibitem
\bibitem{TMVA4}
A.~Hoecker {\em et~al.}, \ifthenelse{\boolean{articletitles}}{\emph{{TMVA 4 --- Toolkit for Multivariate Data Analysis with ROOT. Users Guide.}}, }{}\href{http://arxiv.org/abs/physics/0703039}{{\normalfont\ttfamily arXiv:physics/0703039}}\relax
\mciteBstWouldAddEndPuncttrue
\mciteSetBstMidEndSepPunct{\mcitedefaultmidpunct}
{\mcitedefaultendpunct}{\mcitedefaultseppunct}\relax
\EndOfBibitem
\bibitem{Punzi:2003bu}
G.~Punzi, \ifthenelse{\boolean{articletitles}}{\emph{{Sensitivity of searches for new signals and its optimization}}, }{}eConf \textbf{C030908} (2003) MODT002, \href{http://arxiv.org/abs/physics/0308063}{{\normalfont\ttfamily arXiv:physics/0308063}}\relax
\mciteBstWouldAddEndPuncttrue
\mciteSetBstMidEndSepPunct{\mcitedefaultmidpunct}
{\mcitedefaultendpunct}{\mcitedefaultseppunct}\relax
\EndOfBibitem
\bibitem{JohnsonSU}
N.~L. Johnson, \ifthenelse{\boolean{articletitles}}{\emph{{Systems of frequency curves generated by methods of translation}}, }{}\href{https://doi.org/10.1093/biomet/36.1-2.149}{Biometrika \textbf{36} (1949) 149}\relax
\mciteBstWouldAddEndPuncttrue
\mciteSetBstMidEndSepPunct{\mcitedefaultmidpunct}
{\mcitedefaultendpunct}{\mcitedefaultseppunct}\relax
\EndOfBibitem
\bibitem{Skwarnicki:1986xj}
T.~Skwarnicki, {\em {A study of the radiative cascade transitions between the Upsilon-prime and Upsilon resonances}}, PhD thesis, Institute of Nuclear Physics, Krakow, 1986, {\href{http://inspirehep.net/record/230779/}{DESY-F31-86-02}}\relax
\mciteBstWouldAddEndPuncttrue
\mciteSetBstMidEndSepPunct{\mcitedefaultmidpunct}
{\mcitedefaultendpunct}{\mcitedefaultseppunct}\relax
\EndOfBibitem
\bibitem{ARGUS:1990hfq}
ARGUS collaboration, H.~Albrecht {\em et~al.}, \ifthenelse{\boolean{articletitles}}{\emph{{Search for hadronic $b \to u$ decays}}, }{}\href{https://doi.org/10.1016/0370-2693(90)91293-K}{Phys.\ Lett.\  \textbf{B241} (1990) 278}\relax
\mciteBstWouldAddEndPuncttrue
\mciteSetBstMidEndSepPunct{\mcitedefaultmidpunct}
{\mcitedefaultendpunct}{\mcitedefaultseppunct}\relax
\EndOfBibitem
\bibitem{LHCb-PAPER-2020-046}
LHCb collaboration, R.~Aaij {\em et~al.}, \ifthenelse{\boolean{articletitles}}{\emph{{Precise measurement of the $f_s/f_d$ ratio of fragmentation fractions and of $B^0_s$ decay branching fractions}}, }{}\href{https://doi.org/10.1103/PhysRevD.104.032005}{Phys.\ Rev.\  \textbf{D104} (2021) 032005}, \href{http://arxiv.org/abs/2103.06810}{{\normalfont\ttfamily arXiv:2103.06810}}\relax
\mciteBstWouldAddEndPuncttrue
\mciteSetBstMidEndSepPunct{\mcitedefaultmidpunct}
{\mcitedefaultendpunct}{\mcitedefaultseppunct}\relax
\EndOfBibitem
\bibitem{BaBar:2011vfx}
BaBar collaboration, J.~P. Lees {\em et~al.}, \ifthenelse{\boolean{articletitles}}{\emph{{Amplitude analysis of $B^0\to K^+ \pi^- \pi^0$ and evidence of direct CP violation in $B\to K^* \pi$ decays}}, }{}\href{https://doi.org/10.1103/PhysRevD.83.112010}{Phys.\ Rev.\  \textbf{D83} (2011) 112010}, \href{http://arxiv.org/abs/1105.0125}{{\normalfont\ttfamily arXiv:1105.0125}}\relax
\mciteBstWouldAddEndPuncttrue
\mciteSetBstMidEndSepPunct{\mcitedefaultmidpunct}
{\mcitedefaultendpunct}{\mcitedefaultseppunct}\relax
\EndOfBibitem
\bibitem{Zhou:2023lbc}
S.-H. Zhou, X.-X. Hai, R.-H. Li, and C.-D. Lu, \ifthenelse{\boolean{articletitles}}{\emph{{Analysis of three-body charmless B-meson decays under the factorization-assisted topological-amplitude approach}}, }{}\href{https://doi.org/10.1103/PhysRevD.107.116023}{Phys.\ Rev.\  \textbf{D107} (2023) 116023}, \href{http://arxiv.org/abs/2305.02811}{{\normalfont\ttfamily arXiv:2305.02811}}\relax
\mciteBstWouldAddEndPuncttrue
\mciteSetBstMidEndSepPunct{\mcitedefaultmidpunct}
{\mcitedefaultendpunct}{\mcitedefaultseppunct}\relax
\EndOfBibitem
\bibitem{Wang:2020saq}
W.-F. Wang, J.~Chai, and A.-J. Ma, \ifthenelse{\boolean{articletitles}}{\emph{{Contributions of $K^*_0(1430)$ and $K^*_0(1950)$ in the three-body decays $B\to K\pi h$}}, }{}\href{https://doi.org/10.1007/JHEP03(2020)162}{JHEP \textbf{03} (2020) 162}, \href{http://arxiv.org/abs/2001.00355}{{\normalfont\ttfamily arXiv:2001.00355}}\relax
\mciteBstWouldAddEndPuncttrue
\mciteSetBstMidEndSepPunct{\mcitedefaultmidpunct}
{\mcitedefaultendpunct}{\mcitedefaultseppunct}\relax
\EndOfBibitem
\bibitem{Belle:2009zue}
Belle collaboration, J.-T. Wei {\em et~al.}, \ifthenelse{\boolean{articletitles}}{\emph{{Measurement of the differential branching fraction and forward-backward asymmetry for $B \to K^{(*)}\ell^+\ell^-$}}, }{}\href{https://doi.org/10.1103/PhysRevLett.103.171801}{Phys.\ Rev.\ Lett.\  \textbf{103} (2009) 171801}, \href{http://arxiv.org/abs/0904.0770}{{\normalfont\ttfamily arXiv:0904.0770}}\relax
\mciteBstWouldAddEndPuncttrue
\mciteSetBstMidEndSepPunct{\mcitedefaultmidpunct}
{\mcitedefaultendpunct}{\mcitedefaultseppunct}\relax
\EndOfBibitem
\bibitem{BaBar:2008jdv}
BaBar collaboration, B.~Aubert {\em et~al.}, \ifthenelse{\boolean{articletitles}}{\emph{{Direct CP, lepton flavor and isospin asymmetries in the decays $B \to K^{(*)} \ell^{+} \ell^{-}$}}, }{}\href{https://doi.org/10.1103/PhysRevLett.102.091803}{Phys.\ Rev.\ Lett.\  \textbf{102} (2009) 091803}, \href{http://arxiv.org/abs/0807.4119}{{\normalfont\ttfamily arXiv:0807.4119}}\relax
\mciteBstWouldAddEndPuncttrue
\mciteSetBstMidEndSepPunct{\mcitedefaultmidpunct}
{\mcitedefaultendpunct}{\mcitedefaultseppunct}\relax
\EndOfBibitem
\bibitem{BaBar:2007nlh}
BaBar collaboration, B.~Aubert {\em et~al.}, \ifthenelse{\boolean{articletitles}}{\emph{{Measurements of the branching fractions of $\Bz \to \Kstarz\Kp\Km$, $\Bz\to\Kstarz\pip\Km$, $\Bz\to\Kstarz\Kp\pim$, and $\Bz\to\Kstarz\pip\pim$}}, }{}\href{https://doi.org/10.1103/PhysRevD.76.071104}{Phys.\ Rev.\  \textbf{D76} (2007) 071104}, \href{http://arxiv.org/abs/0708.2543}{{\normalfont\ttfamily arXiv:0708.2543}}\relax
\mciteBstWouldAddEndPuncttrue
\mciteSetBstMidEndSepPunct{\mcitedefaultmidpunct}
{\mcitedefaultendpunct}{\mcitedefaultseppunct}\relax
\EndOfBibitem
\bibitem{LHCb-DP-2022-002}
LHCb collaboration, R.~Aaij {\em et~al.}, \ifthenelse{\boolean{articletitles}}{\emph{{The LHCb Upgrade I}}, }{}\href{https://doi.org/10.1088/1748-0221/19/05/P05065}{{JINST} \textbf{19} (2024) P05065}, \href{http://arxiv.org/abs/2305.10515}{{\normalfont\ttfamily arXiv:2305.10515}}\relax
\mciteBstWouldAddEndPuncttrue
\mciteSetBstMidEndSepPunct{\mcitedefaultmidpunct}
{\mcitedefaultendpunct}{\mcitedefaultseppunct}\relax
\EndOfBibitem
\bibitem{LHCb-PAPER-2025-056-cds}
\url{https://cds.cern.ch/record/2951968}\relax
\mciteBstWouldAddEndPuncttrue
\mciteSetBstMidEndSepPunct{\mcitedefaultmidpunct}
{\mcitedefaultendpunct}{\mcitedefaultseppunct}\relax
\EndOfBibitem
\bibitem{LHCb-PAPER-2024-002}
LHCb collaboration, R.~Aaij {\em et~al.}, \ifthenelse{\boolean{articletitles}}{\emph{{Amplitude analysis of the radiative decay $B^0_s\to K^+K^-\gamma$}}, }{}\href{https://doi.org/10.1007/JHEP08(2024)093}{{JHEP} \textbf{{08}} (2024) {093}}, \href{http://arxiv.org/abs/2406.00235}{{\normalfont\ttfamily arXiv:2406.00235}}\relax
\mciteBstWouldAddEndPuncttrue
\mciteSetBstMidEndSepPunct{\mcitedefaultmidpunct}
{\mcitedefaultendpunct}{\mcitedefaultseppunct}\relax
\EndOfBibitem
\bibitem{LHCb-PAPER-2023-036}
LHCb collaboration, R.~Aaij {\em et~al.}, \ifthenelse{\boolean{articletitles}}{\emph{{Amplitude analysis of the $\Lb \to \proton \Km \gamma$ decay}}, }{}\href{https://doi.org/10.1007/JHEP06(2024)098}{{JHEP} \textbf{06} (2024) 098}, \href{http://arxiv.org/abs/2403.03710}{{\normalfont\ttfamily arXiv:2403.03710}}\relax
\mciteBstWouldAddEndPuncttrue
\mciteSetBstMidEndSepPunct{\mcitedefaultmidpunct}
{\mcitedefaultendpunct}{\mcitedefaultseppunct}\relax
\EndOfBibitem
\end{mcitethebibliography}

\newpage
% LHCb collaboration author list
% Data extracted on December 19th, 2025 at 4:22pm for paper reference LHCb-PAPER-2025-056
\centerline
{\large\bf LHCb collaboration}
\begin
{flushleft}
\small
R.~Aaij$^{38}$\lhcborcid{0000-0003-0533-1952},
A.S.W.~Abdelmotteleb$^{58}$\lhcborcid{0000-0001-7905-0542},
C.~Abellan~Beteta$^{52}$\lhcborcid{0009-0009-0869-6798},
F.~Abudin{\'e}n$^{60}$\lhcborcid{0000-0002-6737-3528},
T.~Ackernley$^{62}$\lhcborcid{0000-0002-5951-3498},
A. A. ~Adefisoye$^{70}$\lhcborcid{0000-0003-2448-1550},
B.~Adeva$^{48}$\lhcborcid{0000-0001-9756-3712},
M.~Adinolfi$^{56}$\lhcborcid{0000-0002-1326-1264},
P.~Adlarson$^{86}$\lhcborcid{0000-0001-6280-3851},
C.~Agapopoulou$^{14}$\lhcborcid{0000-0002-2368-0147}
% ...

{\footnotesize \it

% ...
$^{14}$Universit{\'e} Paris-Saclay, CNRS/IN2P3, IJCLab, Orsay, France\\
% ...
$^{38}$Nikhef National Institute for Subatomic Physics, Amsterdam, Netherlands\\
% ...
$^{48}$Instituto Galego de F{\'\i}sica de Altas Enerx{\'\i}as (IGFAE), Universidade de Santiago de Compostela, Santiago de Compostela, Spain\\
% ...
$^{52}$Physik-Institut, Universit{\"a}t Z{\"u}rich, Z{\"u}rich, Switzerland\\
% ...
$^{56}$H.H. Wills Physics Laboratory, University of Bristol, Bristol, United Kingdom\\
% ...
$^{58}$Department of Physics, University of Warwick, Coventry, United Kingdom\\
% ...
$^{60}$School of Physics and Astronomy, University of Edinburgh, Edinburgh, United Kingdom\\
$^{61}$School of Physics and Astronomy, University of Glasgow, Glasgow, United Kingdom\\
$^{62}$Oliver Lodge Laboratory, University of Liverpool, Liverpool, United Kingdom\\
% ...
$^{70}$Syracuse University, Syracuse, NY, United States\\
% ...
$^{86}$Department of Physics and Astronomy, Uppsala University, Uppsala, Sweden, associated to $^{61}$\\
% ...
\medskip
}
\end{flushleft}

\end{document}